\documentstyle[sprocl,epsfig]{article}
\begin{document}
\title{Physics Beyond the Standard Model
}
\author{R. D. Peccei}
\address{Department of Physics and Astronomy, UCLA, Los Angeles, CA 
90095-1547}
\maketitle
\begin{abstract}

These lectures describe why one believes there is physics beyond the Standard Model and review the expectations of three alternative explanations for the Fermi scale. After examining constraints and hints for beyond the Standard Model physics coming from experiment, I discuss, in turn, dynamical symmetry breaking, supersymmetry and extra compact dimension scenarios associated with  the electroweak breakdown.

\end{abstract}

\section{Introduction}

The Standard Model (SM),~\cite{SM} gives an excellent theoretical
description of the strong and electroweak interactions.  This
theory, which is based on an \break $SU(3)\times SU(2)\times U(1)$ gauge
group, has proven extraordinarily robust.  As shown in Fig. 1, all
data in the electroweak sector to date appears to be in perfect
agreement with the SM predictions, and there are just a few (quite
indirect) hints for physics beyond the SM.  Nevertheless, there are
theoretical aspects of the SM which suggest the need for new physics.
In addition, there are certain open questions within the SM whose
answers can only be found by invoking physics beyond the SM.

In the last year, the observation  by the
SuperKamiokande~\cite{SK} collaboration of neutrino oscillations provided the first experimental
indication that some new physics exists which causes a large splitting
among the leptonic doublets.  While in the quark sector $m_u/m_d\sim
O(1)$, it appears that in the leptonic sector $m_{\nu_\ell}/m_\ell\leq
10^{-8}$.  As we shall see, the most natural explanation for this
phenomena is the existence of a new scale far above the electroweak
scale.

In these Lectures I will try to explore some of the new physics scenarios
which are
motivated by theoretical considerations and try to confront and constrain them
with what we know experimentally, both from the indirect hints coming from the
electroweak sector as well as from the more direct hints 
coming from neutrino oscillations.

\begin{figure}[t]
\center
\epsfig{file=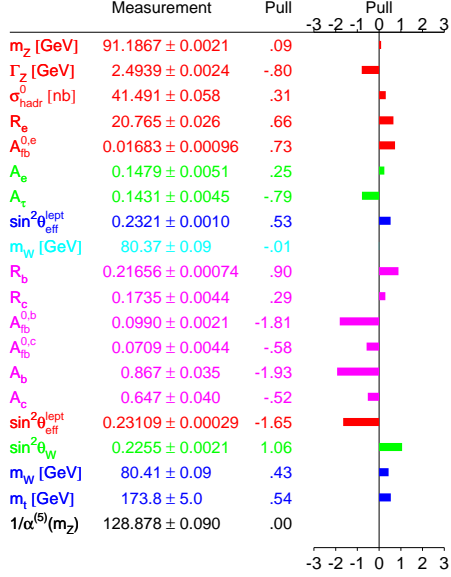,height=3in}
\caption{Standard Model fit, from Ref.2.}
\end{figure}

\section{Theoretical Issues in the Standard Model}

The Standard Model Lagrangian can be written as the sum of four
pieces.  Schematically, one has
\begin{equation}
{\cal{L}}_{\rm SM} = -\sum_f\bar\psi_f\gamma^\mu\frac{1}{i} D_\mu
\psi_f - \frac{1}{4}\sum_i F_i^{\mu\nu}F_{i\mu\nu} +
{\cal{L}}_{\rm SB} + {\cal{L}}_{\rm Yukawa}~.
\end{equation}
The first two terms in the above Lagrangian contain the interactions of the
fermions in the theory with the gauge fields and the 
self-interactions of the gauge fields.  The precision electroweak measurements, 
as well as most QCD tests, essentially have checked these pieces of the SM.
In fact, for the electroweak tests, all that the symmetry-breaking piece,
${\cal{L}}_{\rm SB}$ and the Yukawa piece, ${\cal{L}}_{\rm Yukawa}$, provide are a {\bf renormalizable cut-off} $M_H$ and a {\bf large}
fermion mass $m_t$, respectively.  Of course, ${\cal{L}}_{\rm SB}$ also
allows for the
spontaneous generation of mass for the $W$ and $Z$ bosons.  The masses 
of these excitations are given by the formulas
\begin{equation}
M^2_W = \frac{1}{4} g^2v_F^2~; ~~
M^2_Z = \frac{1}{4} (g^2+g^{\prime 2}) v_F^2
\end{equation}
which involve the $SU(2)[g\equiv g_2]$, and $U(1)[g^\prime]$, coupling constants
as well as a mass scale, $v_F$, arising from ${\cal{L}}_{\rm SB}$.  This
scale---the Fermi scale---is related to the Fermi constant $G_F$ and
sets the scale of the electroweak interactions:
\begin{equation}
v_F = (\sqrt{2} G_F)^{-1/2} \simeq 250~{\rm GeV}~.
\end{equation}

It is important to note that neither the pure gauge field piece of
${\cal{L}}_{\rm SM}$ nor the fermion-gauge piece of this Lagrangian
contain explicit mass terms.  A mass term for the gauge fields
\begin{equation}
{\cal{L}}_{\rm mass}^{\rm Gauge} = -\frac{1}{2} \sum_i m^2_i
A^\mu_iA_{\mu i}
\end{equation}
is forbidden explicitly by the local $SU(3)\times SU(2)\times U(1)$
symmetry.  However, masses for the gauge fields [cf. Eq. (2)] can arise
after the symmetry breakdown $SU(2)\times U(1)\to U(1)_{\rm em}$.
Similarly, fermion mass terms of the form
\begin{equation}
{\cal{L}}_{\rm mass}^{\rm Fermions} = -\sum_f m_f
(\bar\psi_{f{\rm L}}\psi_{f{\rm R}} + \bar\psi_{f{\rm R}}
\psi_{f{\rm L}})
\end{equation}
are forbidden by the $SU(2)\times U(1)$ assignments of the fermions.
This follows since all left-handed fermions are part of $SU(2)$ doublets,
while all right-handed fermions are $SU(2)$ singlets.

Because of these circumstances, mass generation in the Standard Model
is intimately connected to the spontaneous breakdown of the electroweak
$SU(2)\times U(1)$ symmetry.  As a result, not only are the gauge boson masses
$M_W$ and $M_Z$ proportional to the Fermi scale $v_F$, but so are the
masses for all the fermions as well as the mass of the Higgs boson $H$
\begin{equation}
m_f \sim v_F~; ~~~
M_H \sim v_F~.
\end{equation}
The difference between Eq. (2) and Eq. (6) is that, 
in the latter case, the proportionality
constants are {\bf not} known.  Or, better said,
they are related to phenomena we have not yet seen.

There are, however, two important exceptions to the pattern given by
Eq. (6).  First of all, the masses of hadrons are not simply related
to the masses of quarks.  Thus they depend on another scale besides
$v_F$.  This scale, $\Lambda_{\rm QCD}$, is a {\bf dynamical scale}
whose magnitude can be inferred from the running of the $SU(3)$
coupling constant.  A convenient definition is to take $\Lambda_{\rm QCD}$
to be the scale where $\alpha_3(q^2)=g^2_3(q^2)/4\pi$ becomes of
$O(1)$.\footnote{There is no equivalent dynamical scale for the weak
$SU(2)$ group since its coupling becomes strong at scales much below the
scale $v_F$, where the $SU(2)$ group breaks down.}  Then 
$\Lambda_{\rm QCD}$
serves to set the mass scale of the light hadrons which receive the
bulk of their mass from QCD dynamical effects.\footnote{The mass
squared of the pseudoscalar octet is an interesting exception.  Since
these states are quasi-Nambu Goldstone bosons their mass squared is
proportional to the light quark masses.  In fact, one has $m_\pi^2\sim
v_F\Lambda_{\rm QCD}$.\cite{RDPSch}}  
Hadrons containing heavy quarks, on the other
hand, get most of their mass from the mass of the heavy quark.  Thus, less
of their mass depends on QCD dynamics and $\Lambda_{\rm QCD}$.

The second exception to Eq. (6) is provided by neutrinos.  It is clear that for
any particle carrying electromagnetic charge the only allowed mass term
must involve particles and antiparticles, as detailed in Eq.
(5).  Lorentz invariance, however, allows one to write down mass terms
involving two particle fields, or two antiparticle fields.  Such
mass terms, called Majorana mass terms, are allowed for neutrinos.  
In particular,
since the right-handed neutrinos have no $SU(2)\times U(1)$ quantum
numbers, one can write down an $SU(2)\times U(1)$ invariant mass term
for these states of the form
\begin{equation}
{\cal{L}}_{\rm mass}^{\rm Majorana} = -\frac{1}{2}
(\nu_{\rm R}^T M_{\rm R}\tilde C\nu_{\rm R} +
\bar\nu_{\rm R}\tilde c M_{\rm R}^\dagger\bar\nu_{\rm R}^T)~.
\end{equation}
Here $\tilde C$ is a charge conjugation matrix needed for Lorentz
invariance.\cite{RDPMex}  The right-handed neutrino mass matrix $M_{\rm R}$ contains mass scales which are totally {\bf independent} from
$v_F$.  We will return to this point later on in these lectures.

Ignoring these more detailed questions,
one of the principal issues which remains open in the Standard Model
is the nature of the Fermi scale $v_F$.  The role of symmetry
breakdown as a generator of mass scales is familiar in superconductivity.
In that case, the formation of an electron number violating Cooper pair
$\langle e^\uparrow e^\downarrow\rangle$~\cite{BCS} sets up a mass
gap between the normal and the superconducting ground states.  The
Fermi scale $v_F$ plays an analogous role in the electroweak
theory.  It is the scale of the order parameter which is responsible for
the breakdown of $SU(2)\times U(1)$ down to $U(1)_{\rm em}$.  

Although the size of $v_F$ $(v_F\sim 250~{\rm GeV})$ is
known, its precise origin is yet unclear.
Two possibilities have been suggested for the origin of $v_F$:
\begin{description}
\item{i)} The Fermi scale is associated with the vacuum expectation
value (VEV) of some {\bf elementary scalar} field, or fields
$\langle\Phi_i\rangle$.
\item{ii)} The Fermi scale is connected with the formation of some
{\bf dynamical condensates} of fermions of some underlying deeper
theory, $\langle\bar FF\rangle$.
\end{description}
Roughly speaking, the above two alternatives correspond to having
${\cal{L}}_{\rm SB}$ being described either by a weakly coupled theory or by a
strongly coupled theory. 

The nature and origin of the Fermi scale, of course, is not the only
unanswered theoretical question in the SM.  Equally mysterious is the
physics which gives rise to ${\cal{L}}_{\rm Yukawa}$---the piece of the
SM Lagrangian which is responsible for the masses of, and mixing among,
the elementary fermions in the theory.  In contrast to $v_F$, however,
here one does not have directly a scale to associate with this
Lagrangian.  It could well be that the {\bf flavor problem}---the
origin of the fermion masses and of fermion mixing---is the result of physics
operating at scales which are much larger than $v_F$.  Indeed, as we 
will see, trying to generate ${\cal{L}}_{\rm Yukawa}$ itself from
physics at a scale of order $v_F$ is fraught with difficulties.

In view of the above, I will concentrate for now on only on the symmetry
breaking piece of the SM Lagrangian.  
In this context, it proves useful to begin by
examining the simplest example of ${\cal{L}}_{\rm SB}$ in which this
Lagrangian involves just one complex doublet Higgs field:
$\Phi = \left(
\begin{array}{c}
\phi^o \\ \phi^-
\end{array} \right)$:\footnote{Often, one associates the nomenclature
Standard Model to the electroweak theory in which ${\cal{L}}_{\rm SB}$
is precisely given by this simplest option.}
\begin{equation}
{\cal{L}}_{\rm SB} = -(D_\mu\Phi)^\dagger(D^\mu\Phi) -
\lambda\left[\Phi^\dagger\Phi - \frac{1}{2}v_F^2\right]^2~.
\end{equation}
In the above $\lambda$ is an, arbitrary, coupling constant which,
however, must be positive to guarantee a positive definite Hamiltonian.

The Fermi scale $v_F$ enters directly as a scale parameter in the
Higgs potential
\begin{equation}
V = \lambda\left[\Phi^\dagger\Phi - \frac{1}{2}v_F^2\right]^2~.
\end{equation}
The sign in front of the $v_F^2$ term is chosen appropriately to
guarantee that $V$ will be asymmetric, with a minimum at a non-zero
value for $\Phi^\dagger\Phi$.  This fact is what triggers
the breakdown of $SU(2)\times U(1)$ to $U(1)_{\rm em}$, since it forces the field
$\Phi$ to develop a non-zero VEV.\footnote{With only one Higgs doublet
one can always choose $U(1)_{\rm em}$ as the surviving $U(1)$ in the
breakdown.  So the choice $\langle\phi^o\rangle\not= 0$;
$\langle\phi^-\rangle = 0$ is automatic.}
\begin{equation}
\langle\Phi\rangle = \frac{1}{\sqrt{2}} \left(
\begin{array}{c}
v_F \\ 0
\end{array} \right)~.
\end{equation}

Because $v_F$ is an internal scale in the potential $V$, in isolation, it clearly makes no sense to ask what physics
fixes the scale of $v_F$ to be any given particular number.
This question, however, can be asked if one considers the SM in a 
larger context.  For instance, one can imagine that the SM is an
effective theory valid up to some very high cut-off scale $\Lambda$, where
new physics comes in.  An obvious candidate for $\Lambda$ is the
Planck scale $M_P \sim 10^{19}~{\rm GeV}$, the scale associated with
gravity, embodied in Newton's constant $G_N = \frac{1}{M_P^2}$.  In
this broader context then it makes sense to ask what is the relation of
$v_F$ to the cut-off $\Lambda$.  In fact, because the $\lambda\Phi^4$
theory is trivial,~\cite{trivial} with the only consistent theory being
one where $\lambda_{\rm ren} \to 0$, considering the scalar interactions
in ${\cal{L}}_{\rm SB}$ without some high energy cut-off is not sensible.
Let me explain.  

One can readily compute the evolution of the coupling
constant $\lambda$ as a function of $q^2$.  One finds that
$\lambda(q^2)$ evolves in an {\bf opposite} way to the 
way in which the QCD coupling
constant $\alpha_3(q^2)$ evolves, growing as $q^2$ gets larger.  This can be
seen immediately from the Renormalization Group equation (RGE) for $\lambda$  
\begin{equation}
\frac{d\lambda}{d\ln q^2} = +\frac{3}{4\pi^2}\lambda^2 + \ldots
\end{equation}
This equation, in contrast to the QCD case, has a positive rather than a 
negative sign in front of its first term.  As a result, if one solves the above
RGE, including only this first term, one finds a singularity at
large $q^2$ which is a reflection of this growth
\begin{equation}
\lambda(q^2) = \frac{\lambda(\Lambda_o^2)}
{1-\frac{3\lambda(\Lambda_o^2)}{4\pi}\ln\frac{q^2}{\Lambda_o^2}}~.
\end{equation}
This singularity is known as the Landau pole, since Landau was the
first to notice this anomalous kind of behavior.\cite{Landau}

One cannot really trust the location of the Landau pole derived from
Eq. (12)
\begin{equation}
\Lambda_c^2 = \Lambda_o^2\exp\left[\frac{4\pi^2}{3\lambda(\Lambda_o^2)}\right]~,
\end{equation}
since Eq. (12) stops being valid when $\lambda$ gets too large.
When this happens, of course, one should not have neglected the higher order
terms in Eq. (11).  Nevertheless, once the cut-off $\Lambda_c$ is fixed,
one can predict $\lambda(q^2)$ for scales $q^2$ sufficiently below
the cut-off.  Indeed, the $\lambda\Phi^4$ theory is perfectly sensible
as long as one restricts oneself to $q^2\ll \Lambda_c^2$.  If one wants
to push the cut-off to infinity, however, one sees from (13) that
$\lambda(\Lambda_o^2)\to 0$.  This is the statement of triviality,~\cite{trivial}
within this simplified context.

In the case of the SM, one can ``measure" where the cut-off $\Lambda_c$ is in
${\cal{L}}_{\rm SB}$ from the value of the Higgs mass.  Using the
potential (9) one finds that
\begin{equation}
M_H^2 = 2\lambda(M_H^2)v_F^2~.
\end{equation}
Obviously, as long as the Higgs mass is light with respect to $v_F$
the coupling $\lambda$ is small and the cut-off is far away.  Indeed,
using Eqs. (13) and (14) one finds that even if $M_H\sim 200~{\rm GeV}$,
then the cut-off is very large still, of order of the Planck mass
$\Lambda_c\sim M_P$!  So, as long as $M_H$ is that light, or lighter,
the effective theory described by ${\cal{L}}_{\rm SB}$ is very
reliable, and {\bf weakly coupled}, with $\lambda\leq 0.3$.
In these circumstances it is meaningful to ask the question whether the
large hierarchy
\begin{equation}
v_F \ll \Lambda_c
\end{equation}
is a stable condition.  This question, following 't Hooft,~\cite{tH}
is often called the problem of {\bf naturalness}.

If, on the other hand, the Higgs mass is heavy, of order of the
cut-off $(M_H \sim \Lambda_c)$, then it is pretty clear that
${\cal{L}}_{\rm SB}$ as an effective theory stops making sense.
The coupling $\lambda$ is so strong that one cannot separate the 
particle-like excitations from the cut-off itself.  Numerical
investigations on the lattice~\cite{lattice} have indicated that this
occurs when
\begin{equation}
M_H \sim\Lambda_c \sim 700~{\rm GeV}~.
\end{equation}
In this case, it is clear that $\langle\Phi\rangle$, as the order
parameter of the symmetry breakdown, must be replaced by something
else.

Before discussing this latter point,
let me first return to the light Higgs case.  Here one must
worry about the naturalness of having the Fermi scale $v_F$ be
so much smaller that the Planck mass $M_P$, which is clearly a physical
cut-off.  It turns out, in general, that the hierarchy $v_F \ll M_P$
is {\bf not} stable.  This is easy to see since radiative effects in a
theory with a cut-off destabilize any pre-existing hierarchy.  Indeed,
this was 't Hooft's original argument.\cite{tH}  Quantities that are
{\bf not protected} by symmetries suffer quadratic mass shifts.  This
is the case for the Higgs mass.  This mass, schematically, shifts from the 
value given in Eq. (14) to
\begin{equation}
M_H^2 = 2\lambda v_F^2 + \alpha \Lambda_c^2~.
\end{equation}
It follows from Eq. (17) that
if $\Lambda_c\sim M_P \gg v_F$, the Higgs bosons cannot remain
light.  Or, saying it another way, if one wants the Higgs to remain
light, one needs an enormous amount of fine tuning of parameters to
guarantee that, in the end, it remains a light excitation.  This kind
of fine-tuning is really unacceptable, so one is invited to look for some
protective symmetry to guarantee that the hierarchy $v_F\ll M_P$
is stable.\footnote{Note that a stable hierarchy 
$v_F \ll M_P$ does not explain why one has such a hierarchy to begin
with.  This is a much harder question to answer.}

Such a protective symmetry exists---it is supersymmetry (SUSY).\cite{SUSY}
SUSY is a boson-fermion symmetry in which bosonic degrees of freedom
are paired with fermionic degrees of freedom.  If supersymmetry is
exact then the masses of the fermions and of their bosonic partners
are the same.  In a supersymmetric version of the Standard Model
all quadratic divergences cancel.  Thus parameters like the Higgs boson mass
will not be sensitive to a high energy cut-off.  Roughly speaking, via
supersymmetry, the Higgs boson mass is kept light naturally since its
fermionic partner has a mass which is protected by a chiral symmetry
and is of $O(v_F)$.

Because one has not seen any of the SUSY partners of the states in the
SM yet, it is clear that if a supersymmetric extension of the SM exists
then the associated supersymmetry must be broken.  Remarkably, even if
SUSY is broken the naturalness problem in the SM is resolved, provided
that the splitting between the fermion-boson SUSY partners is itself
of $O(v_F)$.  For example, the quadratic divergence of the Higgs mass
due to a $W$-loop is moderated into only a logarithmic divergence by the
presence of a loop of Winos, the spin-1/2 SUSY partners of the $W$
bosons.  Schematically, in the SUSY case, Eq. (17) gets replaced by
\begin{equation}
M_H^2 = 2\lambda v_F^2 + \alpha(\tilde M_W^2-M_W^2)
\ln \Lambda_c/v_F~.
\end{equation}
So, as long as the masses of the SUSY partners (denoted by a tilde)
are themselves not split away by much more than $v_F$, radiative
corrections will not destabilize the hierarchy $v_F\ll\Lambda_c$.

Let me recapitulate.  Theoretical considerations regarding the nature
of the Fermi scale have brought us to consider two alternatives for new
physics associated with the $SU(2)\times U(1)\to U(1)_{\rm em}$ 
breakdown and ${\cal{L}}_{\rm SB}$:
\begin{description}
\item{i)} ${\cal{L}}_{\rm SB}$ is the Lagrangian of some elementary scalar
fields interacting together via an asymmetric potential, whose minimum
is set by the Fermi scale $v_F$.  The presence of non-vanishing
VEVs triggers the electroweak breakdown.  However, to guarantee the
naturalness of the hierarchy $v_F\ll M_P$, both ${\cal{L}}_{\rm SB}$
and the whole Standard Model Lagrangian must be augmented by other
fields and interactions so as to enable the theory (at least
approximately) to be supersymmetric.  Obviously, if this alternative is
true, there is plenty of new physics to be discovered,
since all particles have
superpartners of mass $\tilde m \simeq m + O(v_F)$.
\item{ii)} The symmetry breaking sector of the SM has itself a dynamical
cut-off of $O(v_F)$.  In this case, it makes no sense to describe
${\cal{L}}_{\rm SB}$ in terms of strongly coupled scalar fields.
Rather, ${\cal{L}}_{\rm SB}$ describes a dynamical theory of some new
strongly interacting fermions $F$, whose condensates cause the
$SU(2)\times U(1)\to U(1)_{\rm em}$ breakdown.  The strong interactions
which form the condensates $\langle\bar FF\rangle\sim v_F^3$ also
identify the Fermi scale as the dynamical scale of the underlying
theory, very much analogous to $\Lambda_{\rm QCD}$.  If this alternative
turns out to be true, then one expects in the future to see lots of new
physics, connected with these new strong interactions, when one probes
them at energies of $O(v_F)$.
\end{description}

In the past year, a third very speculative alternative has been
suggested besides the two possibilities above.\cite{ADD}  This alternative
is based on the idea that, perhaps, in nature there could exist some
extra ``largish" compact dimensions of size R.~\cite{Antoniadis}
In such theories, the fundamental scale of gravity in $(d+4)$-dimensions
could well be quite different than the Planck scale.  In particular, it
may well be that
\begin{equation}
[M_P]_{d+4} \sim v_F~.
\end{equation}
That is, at short distances $(r<R)$ the scale of gravity could well be
different than the Planck scale, the usual scale of 4-dimensional
gravity valid at large distances $(r>R)$.  Indeed, this 
short-distance scale could be
identical to the Fermi scale.  The relationship between
these two gravity scales depends both on $R$ and on the number of
extra dimensions $d$:~\cite{ADD}
\begin{equation}
M_P = ([M_P]_{d+4})^{\frac{d+2}{2}} R^{\frac{d}{2}} \sim
v_F[v_FR]^{\frac{d}{2}}
\end{equation}
where the second (approximate) equality holds if Eq. (19) holds.

Obviously, if Eq. (19) were to be true, then there is no naturalness
issue---the
Fermi scale is the scale of gravity in the ``true" 
extra-dimensional theory!  From
Eq. (20) it follows that, if $d=2$, then the scale of the compact
dimensions needed is quite large $R\sim 10^{-1}~{\rm cm}$!  On the 
other hand, if $d=6$, as string theory suggests, then 
$R\sim 10^{-13}~{\rm cm}$.  These distances are small enough that 
perhaps one would not have noticed the modifications implied for the
gravitational potential at distances $r<R$.

Although these theories do not suffer from any naturalness problem,
and thus are perfectly consistent with a single weakly-coupled Higgs
field, they do predict the existence of other phenomena beyond the
SM.  In particular, if this alternative is correct, one would expect
copious production of gravitons at energies of order $\sqrt{s}\sim
v_F$, as one begins to excite the compact dimensions.  Thus, also
here there is spectacular new physics to find!

In these Lectures, I will try to illustrate some of the consequences
of all the three alternatives for $v_F$ alluded to above.  In
addition, to try to divine which of the above ideas is most likely to be
correct, 
I want to explore in some depth some of the points which come 
from experiment suggesting possible traces of physics beyond the
Standard Model. In the next section, I will try to describe
in more detail what these hints of physics beyond the Standard Model 
are and what is their likely origin.

\section{Constraints and Hints for Beyond the Standard Model Physics}

There are four different experimental inputs which help shed some light on
possible physics beyond the Standard Model.  I will discuss them in turn.

\subsection{Implications of Standard Model Fits}

One of the strongest constraints on physics beyond the SM is that the SM
gives an excellent fit to the data, as we already illustrated in Fig. 1.  In
practice, since all fermions but the top are quite light compared to the
scale of the $W$ and $Z$-bosons, all quantities in the SM are specified as functions
of 5 parameters: $g$; $g^\prime$; $v_F$; $M_H$; and $m_t$.  It proves 
convenient to trade the first three of these for another triplet of 
quantities in the SM which are better measured:  $\alpha$; $M_Z$; and
$G_F$.  This trade-off has become the common practice in the field.  Once
one has adopted a set of {\bf standard parameters} then all physical
measurable quantities can be expressed as a function of this ``standard set".
For example, the $W$-mass in the SM is given as a function of these
parameters as:
\begin{equation}
\left. M_W\right|_{\rm SM} = M_W(\alpha;M_Z;G_F;m_t;M_H)~.
\end{equation}

Because $\alpha$, $M_Z$, and $G_F$, as well as $m_t$\footnote{The top mass
is quite accurately determined now.  The combined value obtained by the CDF
and DO collaborations fixes $m_t$ to better than 3\%:  $m_t =
(174.8 \pm 5.0)$ GeV.\cite{Tevatron}} are rather accurately known, all SM
fits essentially serve to constrain only {\bf one} unknown--the Higgs mass
$M_H$.  This constraint, however, is not particularly strong because all
radiative effects depends on $M_H$ only logarithmically.  That is, radiative
corrections give contributions of $O\left(\frac{\alpha}{\pi} \ln M_H/M_Z\right)$.

The result of the SM fit of all precision data gives for the Higgs mass the
result:~\cite{fit}
\begin{equation}
M_H = \left(76^{\scriptstyle +85}_{\scriptstyle -47}\right)~{\rm GeV}
\end{equation}
or
\begin{equation}
M_H < 262~{\rm GeV}~~(95\% ~{\rm C.L.})~.
\end{equation}
These results for the Higgs mass are compatible with limits on $M_H$ coming
from direct searches for the Higgs boson in the process $e^+e^-\to ZH$ at
LEP 200.  The limit presented at the 1998 ICHEP in Vancouver
was~\cite{Vancouver} 

\begin{equation}
M_H > 89.8~{\rm GeV}~~~(95\%~{\rm C.L.})~.
\end{equation}
However, preliminary results presented at the 1999 Winter Conferences have
raised this bound to the neighborhood of 95 GeV.

It is particularly gratifying that the SM fits indicate the need for a light
Higgs boson, since this ``solution" 
is what is internally consistent.  Let me illustrate how this emerges, for
example, from studies of the $Z$-leptonic vertex. 
The axial coupling of the $Z$
\begin{equation}
{\cal{L}}_{\rm eff} = \frac{eZ^\mu}{2\cos\theta_W\sin\theta_W}
\bar e[\gamma_\mu g_V-g_A\gamma_\mu\gamma_5]e
\end{equation}
gets modified by radiative corrections to
\begin{equation}
g_A^2 = \frac{1}{4}[1+\Delta\rho]~.
\end{equation}
The shift in the $\rho$-parameter, $\Delta\rho$, gets its principal contribution
from $m_t$.  However, it has also a (weak) dependence on
$M_H$.\cite{Alta}
\begin{equation}
\left.\Delta\rho\right|_{\rm Higgs} = -\frac{3G_FM^2_W}{4\pi^2\sqrt{2}}
\tan^2\theta_W\ln\frac{M_H}{M_Z} + \ldots \simeq
-10^{-3}\ln\frac{M_H}{M_Z}~.
\end{equation}
The SM fit gives~\cite{fit}
\begin{equation}
\Delta\rho = (3.7\pm 1.1)\times 10^{-3}~,
\end{equation}
with the Higgs contribution giving, for $M_H = 300$ GeV,
$\Delta\rho = -1\times 10^{-3}$.  Obviously, if $M_H$ were to be very
large, the Higgs contribution could have even changed the sign of
$\Delta\rho$.  The value emerging from the SM fit instead is perfectly
compatible with having a rather light Higgs mass.  In fact, one nice way to
summarize the result of the SM fit is that, approximately, this fit constrains
\begin{equation}
\ln \frac{M_H}{M_Z} \leq 1~.
\end{equation}
That is, there are {\bf no} large logarithms associated with the symmetry
breaking sector.

I should remark that a good SM fit does not necessarily exclude possible
extensions of the SM involving either new particles or new interactions,
provided that these new particles and/or interactions give only small effects.
Typically, the effects of new physics are small if the excitations associated 
with this new physics have mass scales several times the $W$-mass.

One can quantify the above discussion in a more precise way by introducing
a general parametrization for the vacuum polarization tensors of the gauge
bosons and the $Zb\bar b$ vertex. These are the places where the dominant
electroweak radiative corrections occur and therefore are the quantities
which are probably the most sensitive new physics.\cite{epsilon}  I do not
want to enter into a full discussion of this procedure here since it is already
well explained in the literature.\cite{epsilon}  However, I want to
talk about one example, connected to modifications of the gauge fields vacuum
polarization tensors, because precision electroweak data serves to provide a strong constraint
on dynamical symmetry breaking theories-excluding theories which
are QCD-like.

There are four distinct vacuum polarization contributions
$\Sigma_{AB}(q^2)$, where the pairs $AB = \{ZZ,WW,\gamma\gamma,
\gamma Z\}$.  For sufficiently low values of the momentum transfer
$q^2~(q^2\simeq M^2_W)$ it obviously suffices to expand 
$\Sigma_{AB}(q^2)$ only up to $O(q^2)$.  Thus, approximately,
one needs to consider 8 different parameters associated with these
contributions:
\begin{equation}
\Sigma_{AB}(q^2) = \Sigma_{AB}(0) + q^2\Sigma_{AB}^\prime(0) + \ldots~,
\end{equation}
with the remaining corrections being terms of $O\left(\frac{q^4}{\Lambda^2}\right)$ with $\Lambda$ being the scale of the new
physics.  In fact, there are not really 8 independent parameters since
electromagnetic gauge invariance requires that
\begin{equation}
\Sigma_{\gamma\gamma}(0) = \Sigma_{\gamma Z}(0) = 0~.
\end{equation}
Of the 6 remaining parameters one can fix 3 combinations of coefficients in
terms of $G_F,~\alpha$ and $M_Z$.  Hence, in a most general analysis, the
gauge field vacuum polarization tensors (for $q^2\stackrel{<}{_{\scriptstyle
\sim}} M_W^2$) only involve 3
arbitrary parameters.  The usual choice,\cite{epsilon} is to have one of
these contain the main quadratic $m_t$-dependence, leaving the other two
essentially independent of $m_t$.

I will proceed with my discussion in terms of the parametrization of 
Altarelli and Barbieri,\cite{epsilon} where these three parameters are chosen to
be
\begin{eqnarray}
\epsilon_1 &=& \Delta\rho = \left[\frac{\Sigma_{ZZ}(0)}{M^2_Z} -
\frac{\Sigma_{WW}(0)}{M^2_W}\right] \stackrel{\rm SM}{=}
\frac{3G_Fm_t^2}{8\pi^2\sqrt{2}}-\frac{3G_FM^2_W}{4\pi^2\sqrt{2}}
\tan^2\theta_W\ln\frac{M_H}{M_Z} \ldots \\
\epsilon_2 &=& \left[\Sigma^\prime_{11}(0)-\Sigma^\prime_{33}(0)\right]
\stackrel{\rm SM}{=} -\frac{G_FM^2_W}{2\pi^2\sqrt{2}}\ln\frac{m_t}{M_Z}
+ \ldots \\
\epsilon_3 &=& \left[\Sigma^\prime_{3\gamma}(0) - \Sigma^\prime_{3Z}(0)\right]
\stackrel{\rm SM}{=} \frac{G_FM^2_W}{12\pi^2\sqrt{2}}
\ln\frac{M_H}{M_Z} - \frac{G_FM^2_W}{6\pi^2\sqrt{2}}\ln\frac{m_t}{M_Z}
+ \ldots
\end{eqnarray}
Here, as usual, $Z = 3-\sin^2\theta_W\gamma$ and $W^\pm = {1\over \sqrt{2}}
[1\mp i2]$. In the above, I 
have displayed also the leading dependence on $m_t$ and $M_H$ of
the $\epsilon_i$ in the SM.

The interesting parameter here is $\epsilon_3$, whose experimental value 
turns out to be~\cite{fit}
\begin{equation}
\epsilon_3 = (3.9\pm 1.1)\times 10^{-3}~.
\end{equation}
One can estimate $\epsilon_3$ in a dynamical symmetry breaking theory,
if one assumes that the spectrum of such a theory, and its dynamics, is
QCD-like.\cite{Peskin}  From its definition, one sees that $\epsilon_3$
involves the difference between the spectral functions of vector and
axial vector currents
\begin{equation}
\epsilon_3 = \frac{1}{2}\left[\Sigma^\prime_{VV}(0) - \Sigma^\prime_{AA}(0)\right]~.
\end{equation}
This difference has two components in a dynamical symmetry breaking theory.
There is a contribution from a heavy Higgs boson $(M_H\sim {\rm TeV})$
characteristic of such theories, plus a term detailing the differences between the vector
and axial vector spectral functions. 
This second component reflects the resonances with these
quantum numbers in the spectrum of the 
underlying theory which gives rise to the symmetry
breakdown.  The first piece is readily estimated from the SM expression,
using $M_H\sim {\rm TeV}$.  The second piece, in a QCD-like theory, can
be deduced by analogy to QCD, modulo some counting factors associated with
the type of underlying theory one is considering.  One finds~\cite{Peskin}
\begin{eqnarray}
\epsilon_3 &=& \left.\epsilon_3\right|_{M_H\simeq 1~{\rm TeV}} +
\frac{\alpha}{12\pi\sin^2\theta_W}\int^\infty_0 \frac{ds}{s}
[R_V(s)-R_A(s)] \nonumber \\
&=& 6.65\times 10^{-3} + N_D\left(\frac{N_{TC}}{4}\right)
\left[\frac{2\alpha\pi}{\sin^2\theta_W}\frac{f_\pi^2}{m_\rho^2}\right] \nonumber \\
&=& \left[6.65\pm 3.4 N_D\left(\frac{N_{TC}}{4}\right)\right]
\times 10^{-3}~.
\end{eqnarray}
The second line above follows if the underlying theory is QCD-like, so that
the resonance spectrum is saturated by $\rho$-like and $A_1$-like, resonances.
Here $N_D$ is the number of doublets entering in the underlying theory
and $N_{TC}$ is the number of ``Technicolors" in this theory.\footnote{For
QCD, of course, $N_D=1$ and $N_{TC}=3$.}  Using Eq. (34) and using
$N_{TC}=4$, as is usually assumed, one sees that
\begin{equation}
\epsilon_3 = \left\{ \begin{array}{ll}
10.05\times 10^{-3} & N_D=1 \\
20.25\times 10^{-3} & N_D=4
\end{array}
\right.
\end{equation}
These values for $\epsilon_3$ are, respectively, 
$5.5\sigma$ and $15\sigma$ away from the best fit value of $\epsilon_3$,
obtained from fitting all the high precision electroweak data.  Obviously,
one cannot countenance anymore a dynamical symmetry breaking theory which
is QCD-like!

Provided the superpartners are not too light,
nothing as disastrous occurs instead if one considers a supersymmetric
extension of the SM.  
Fig. 2, taken from a recent analysis of Altarelli, Barbieri, and Caravaglios,\cite{ABC} shows a typical fit, scanning over a range of
parameters in the MSSM---the minimal supersymmetric extension of the SM.
Although the MSSM can improve the $\chi^2$ of the fit over that for
the SM (which is already very good!), these improvements are small.  In
effect, the MSSM radiative corrections fits are all slightly better than
the SM fits.  This is not surprising, since these latter
fits contain more parameters.  Interestingly, however, these fits do not
provide better bounds on sparticles than the bounds obtained by direct
searches.  Of course, for certain cases there are constraints.  For
instance, there cannot be too large a stop-sbottom splitting because such
a splitting would give too large a value for $\epsilon_1=\Delta\rho$.

\begin{figure}[t]
\center
\epsfig{file=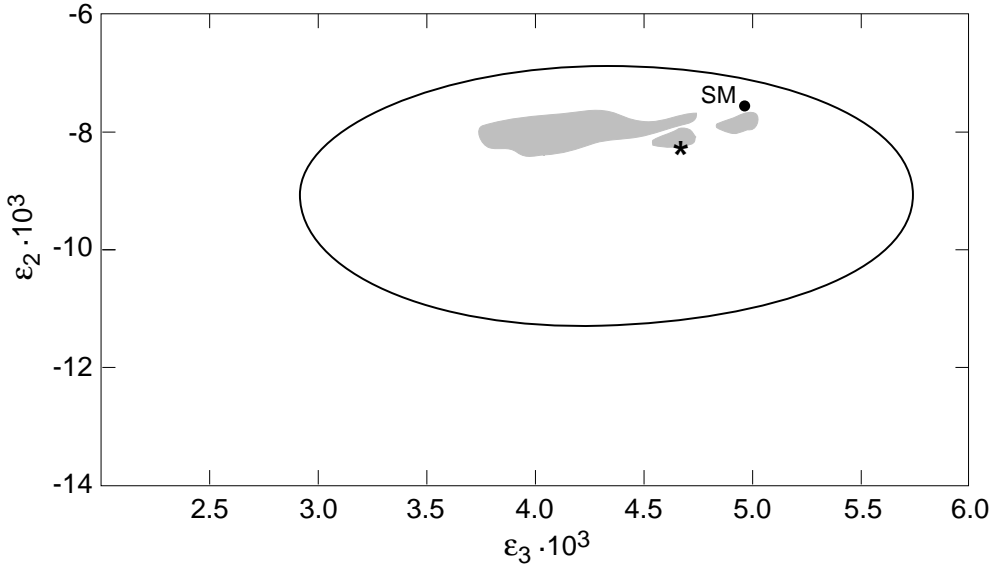,height=3in}
\caption{Comparison of SM and MSSM fits in the $\epsilon_2-\epsilon_3$ plane, from Ref.20. The ellipse is the 1-$\sigma$ range determined by the data. The shaded region is the result of a scan over a range of SUSY parameters, with the star marking the lowest $\chi^2$ point.}
\end{figure}

\subsection{Hints of Unification}

Although the SM coupling constants are very different at energies of order
of the Fermi scale,\footnote{One has, for instance, $\alpha_3(M^2_Z)\simeq
0.12$, while $\alpha_2(M^2_Z)=\alpha(M^2_Z)/\sin^2\theta_W
\simeq 0.034$.} these couplings
can become comparable at very high energies because they evolve differently
with $q^2$.~\cite{GQW}  Indeed, it is quite possible that the SM couplings
unify into a single coupling at high energy, reflecting an underlying
Grand Unified Theory (GUT) which breaks down to the SM at a high scale.
If $G$ denotes the GUT group, then one can imagine the sequence of spontaneous
breakings
\begin{equation}
G\stackrel{M_X}{\longrightarrow} SU(3)\times SU(2)\times U(1)
\stackrel{v_F}{\longrightarrow} SU(3)\times U(1)_{\rm em}~,
\end{equation}
with $M_X\gg v_F$.

To test this assumption one can compute the evolution of the SM coupling
constants using the Renormalization Group Equations (RGE) and see if,
indeed, these coupling constants unify.  To leading order, the evolution of
each coupling constant can be evaluated separately from the others, since
they decouple from each other:
\begin{equation}
\frac{d\alpha_i(\mu^2)}{d\ln\mu^2} = -\frac{b_i}{4\pi}\alpha_i^2(\mu^2)~.
\end{equation}
These equations imply a logarithmic change for the inverse couplings
\begin{equation}
\alpha_i^{-1}(q^2) = \alpha_i^{-1}(M^2) +
\frac{b_i}{4\pi} \ln \frac{q^2}{M^2}~.
\end{equation}

The rate of change of the coupling constants with energy is governed by the
coefficients $b_i$ which enter in the RGE.  In turn, these coefficients
depend on the {\bf matter content} of the theory---which 
matter states are ``active" at the 
scale one is probing.  In general, one has~\cite{Slansky}
\begin{equation}
b_i = \sum_{\rm states}\left\{\frac{11}{6} \ell_i^{\rm vector} -
\frac{1}{3}\ell_i^{\rm chiral~fermion} - \frac{1}{12}\ell_i^{\rm real~scalar}\right\}
\end{equation}
with the $\ell_i$ being group theoretic factors.  For $SU(N)$ groups~\cite{Slansky} $[\ell_i]_{\rm Adjoint} = 2N$, while for fields transforming according to the fundamental representation, 
$[\ell_i]_{\rm Fundamental}=1$.  For $U(1)$ groups, $\ell_i=2Q^2$, where
$Q$ is a ``property normalized" charge.  That is, a charge which allows the
possibility of unifying the $U(1)$ groups with the other non-Abelian groups.
Let me explain this last point further.

For non-Abelian groups, the generators in the fundamental representation
are conventionally normalized~\cite{Slansky} so that
\begin{equation}
{\rm Tr}~t_at_b = \frac{1}{2}\delta_{ab}~.
\end{equation}
For Abelian groups one can always rescale the charge.  Thus no similar
convention exists for this case.  
For example, for the electroweak group, instead of the
usual hypercharge $Y$, one can define a new charge $Q$ related to $Y$ by
a constant:
\begin{equation}
Y = \xi Q~.
\end{equation}
Obviously, the conventional hypercharge coupling can be turned into a $Q$-coupling,
by rescaling the $U(1)$ coupling constant:
\begin{equation}
g^\prime Y = g^\prime\xi Q = g_1Q~.
\end{equation}
For unification, one wants a $U(1)$ charge that is normalized in the same
way as the non-Abelian generators, when one sweeps over all quarks and
leptons
\begin{equation}
\sum_{q+\ell} {\rm Tr}~ t_a^2 = \sum_{q+\ell}{\rm Tr}~ Q^2 \equiv
\frac{1}{\xi^2}\sum_{q+\ell}{\rm Tr}~ Y^2~.
\end{equation}
Because $\displaystyle{\sum_{q+\ell}}{\rm Tr}~t_a^2=2$, while
$\displaystyle{\sum_{q+\ell}}{\rm Tr}~Y^2= 10/3$ it follows that
$\xi=\sqrt{5/3}$.  So,
\begin{equation}
Q=\sqrt{\frac{3}{5}}Y~; ~~~ g_1 = \sqrt{\frac{5}{3}} g^\prime~.
\end{equation}

Using Eq. (42), it is straightforward to compute the coefficients $b_i$
in the SM.  For example, for the QCD coupling, one finds
\begin{equation}
b_3 = \left\{\frac{11}{6}\cdot 6 - \frac{1}{3} \cdot 12\right\} = 7~,
\end{equation}
where the first factor above is the contribution of the gluonic degree of
freedom and the second factor above comes from the 6 species of left-handed
quarks plus the 6 species of right-handed quarks.  Obviously, if there were 
to be supersymmetric matter, all the coefficients $b_i$ would be modified
at scales at or above where this matter starts to be produced.  For
example, in the supersymmetric QCD case, the gluons are now accompanied by
spin 1/2 gluinos (which are chiral fermions) and each of the quarks of a given
helicity has two real spin zero squark partners.  For SUSY QCD then, the
coefficient $b_3$ becomes
\begin{equation}
b_3 = \left\{\frac{11}{6}\cdot 6 - \frac{1}{3}\cdot 12 - \frac{1}{3}\cdot 6 -
\frac{1}{12}\cdot 24 \right\} = 3~.
\end{equation}
Table 1 compares the predictions for the coefficients $b_i$ of the SM and the
SUSY extension.\footnote{Note that the SUSY SM has, by necessity, 2 Higgs
doublets, while the SM is assumed to have only 1 Higgs doublet.}

\begin{table}
\caption{Coefficients $b_i$ in the SM and in the SUSY SM}
{\begin{center}
\begin{tabular}{|c|c|c|} \hline
Coefficient & SM & SUSY SM \\  \hline
$b_1$ & $-41/10$ & $-33/5$ \\
$b_2$ & $+19/6$ & $-1$ \\
$b_3$ & $+7$ & $+3$ \\ \hline
\end{tabular}
\end{center}}
\end{table}

Using the result of this table, along with input data for $\alpha_i(M^2_Z)$~\cite{LEPEW}, one can compute the evolution of the coupling
constants in both models.~\cite{Amaldi}  As is shown in Fig. 3, for the SM
there is a {\bf near unification} of the couplings around $M_X \simeq
10^{15}$ GeV.  However, rather remarkably, in the supersymmetric extension of
the SM, the presence of the SUSY matter, by altering the evolution,
appears to give a {\bf true unification} of the coupling constants at
$M_X\simeq 10^{16}$ GeV.

\begin{figure}[t]
\center
\epsfig{file=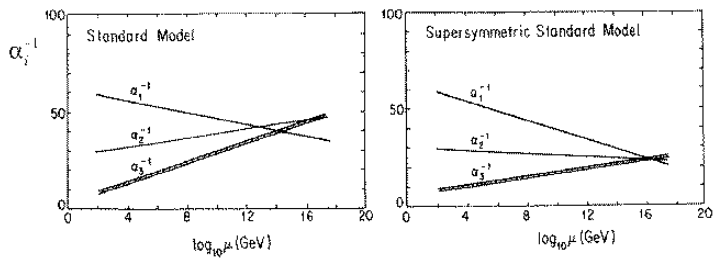,width=0.9 \textwidth}
\caption{Evolution of couplings without and with SUSY matter.}
\end{figure}

The unification of the couplings in the SUSY SM case is quite spectacular.
However, {\it per se}, this is only suggestive.  It is not either a
``proof" that a low energy supersymmetry exists, nor does it mean that there exists
some high energy GUT!  The proof of the former really requires the discovery
of the predicted SUSY partners, while for GUTs one must find typical
phenomena which are associated with these theories--like proton decay.  This
said, however, one can gather additional ammunition in favor of
this picture from some of the properties of the top quark.  I turn to this
point next.

\subsection{Implications of a Large Top Mass}

Of all the quarks and leptons, only top has a mass which is of order of the
Fermi scale, $v_F \simeq 250$ GeV.  In this sense, top is unique among all the
fundamental particles, since it has a mass whose value is basically set by the
value of the order parameter responsible for the electroweak breakdown:
$m_t\sim v_F$.  All other elementary excitations are related to $v_F$ by
constants which are much less than unity.

If one is permitted a perturbative analysis, having a large top mass, in
turn, gives further constraints.  This is particularly true for the case of the 
SM, where a large top mass influences what Higgs masses are allowed.
However, interesting consequences also arise in the (theoretically more
pristine) SUSY SM.  In both cases, rather than dealing with the ``physical"
top quark mass determined experimentally by CDF and DO:~\cite{Tevatron}
\begin{equation}
m_t = (173.8\pm 5.0)~{\rm GeV}~,
\end{equation}
it is more convenient to consider instead the running mass
\begin{equation}
m_t(m_t) \simeq \frac{m_t}{1+\frac{4}{3\pi}\alpha_3(m_t^2)} =
(165\pm 5)~{\rm GeV}~.
\end{equation}
This is because $m_t(m_t)$ is directly related to the diagonal couplings of
the top to the VEV of 
the Higgs boson, $H_u$:\footnote{In the SM there is only one
Higgs boson, so the subscript $u$ is unnecessary.  For the SUSY SM,
$H_u$ is the Higgs field which couples to the right-handed up-quarks
(while $H_d$ couples to the right-handed down-quarks).}
\begin{equation}
m_t(m_t) = \lambda_t(m_t)\langle H_u\rangle~.
\end{equation}

The Yukawa coupling $\lambda_t$ also obeys a RGE.  Keeping only the dominant
3rd generation couplings, this equation reads:~\cite{poles}
\begin{equation}
\frac{d\lambda_t(\mu)}{d\ln \mu^2} = \frac{1}{32\pi^2}
\left[a_t\lambda_t^2(\mu)+a_b\lambda_b^2(\mu)+a_\tau\lambda_\tau^2(\mu) -
4\pi c_i\alpha_i(\mu^2)\right]\lambda_t(\mu)~.
\end{equation}
Here $\lambda_b$ and $\lambda_\tau$ are, respectively, the Yukawa couplings
of the $b$-quark and the $\tau$-lepton to the (corresponding) Higgs
field.  The coefficients $a_t,a_b,a_\tau$ and $c_i$ in Eq. (53) again
depend on the matter content of the theory.  Table 2 details them both for
the SM and its SUSY extension.
\begin{table}
\caption{Coefficients entering the RGE for $\lambda_t$ for the SM and
the SUSY SM}
{\begin{center}
\begin{tabular}{|c|c|c|} \hline
Coefficient & SM & SUSY SM \\ \hline
$a_t$ & $9/2$ & $6$ \\
$a_b$ & $3/2$ & $1$ \\
$a_\tau$ & $1$ & $0$ \\ 
$c_1$ & $17/20$ & $17/5$ \\
$c_2$ & $9/4$ & $3$ \\
$c_3$ & $8$ & $16/3$ \\ \hline
\end{tabular}
\end{center}}
\end{table}

Because the coefficient $a_t>0$, it follows that also the top coupling
$\lambda_t(\mu)$ will have a Landau pole at large values of the scale $\mu$.
Of course, just as  
for the case of the Higgs coupling $\lambda$ discussed earlier,
the location of this singularity is not to be trusted exactly since Eq. (53)
breaks down in its vicinity.  Nevertheless, there are significant
differences between where the Landau pole for $\lambda_t$ is in the SM and
where it is in the SUSY SM.  

\subsubsection{SM Case}

Because one assumes that there is only one Higgs boson in the Standard Model,
it follows that $\langle H_u\rangle = \frac{1}{\sqrt{2}} v_F\simeq 174$ GeV.
This implies, in turn, a precise value for $\lambda_t(m_t)$ from Eq. (52):
\begin{equation}
\lambda_t(m_t) = 0.95 \pm 0.03~.
\end{equation}
Further, since $m_t\gg m_b,m_\tau$ and $c_3\alpha_3\gg c_2\alpha_2,c_1\alpha_1$,
to a good approximation the RGE (53) reduces to
\begin{equation}
\frac{d\lambda_t(\mu)}{d\ln\mu^2} = \frac{1}{32\pi^2}
\left[\frac{9}{2}\lambda^2_t(\mu) - 32\pi\alpha_3(\mu^2)\right]\lambda_t(\mu)~.
\end{equation}
Using $\alpha_3(m_t^2)\simeq 0.118$, one sees that the above square 
bracket is {\bf negative} at $\mu = m_t$.  Thus in the SM, $\lambda_t(\mu)$
{\bf decreases} as $\mu$ increases above $m_t$, at least temporarily.
However, for very large $\mu$, eventually $\lambda_t(\mu)$ will begin
growing and eventually it will diverge at some scale--the Landau pole.

Eq. (55) and its companion for $\alpha_3(\mu^2)$, Eq. (40), can be solved
in closed form.~\cite{poles}  One finds for $\lambda_t(\mu^2)$ the 
expression
\begin{equation}
\lambda_t^2(\mu) = \frac{\eta(\mu)\lambda_t^2(m_t)}
{\left[1-\frac{9}{16\pi^2}\lambda_t^2(m_t)I(\mu)\right]}~,
\end{equation}
where the functions $\eta(\mu)$ and $I(\mu)$ contain information on the
running of the strong coupling constant:
\begin{eqnarray}
\eta(\mu) &=& \left[\frac{\alpha_3(m_t^2)}{\alpha_3(\mu^2)}\right]^{-c_3/b_3}
= \left[\frac{\alpha_3(\mu^2)}{\alpha_3(m_t^2)}\right]^{8/7} \nonumber \\
I(\mu) &=& \int^{\ln\mu}_{\ln m_t} d\ln\mu^\prime \eta(\mu^\prime) 
\end{eqnarray}
Using Eq. (56) it is easy to check that the Yukawa coupling $\lambda_t(\mu)$
decreases well beyond the Planck scale, with $\lambda_t(M_{\rm P})\simeq
0.65 < \lambda_t(m_t)$, so that the top sector is perturbative throughout
the region of interest.  In fact, $\lambda_t(\mu)$ does not begin to get
large until $\mu\sim 10^{30}$ GeV, with the Landau pole occuring around
$\mu = 10^{32}$ GeV---scales well beyond the Planck scale.

Even though the top Yukawa coupling is below unity for $\mu < M_{\rm P}$,
this coupling is large enough to affect the Higgs sector of the Standard
Model.  Our discussion of the Higgs self-coupling $\lambda$ in Section II
was based on the RGE in which {\bf only} terms involving $\lambda$ were
retained.  In fact, at higher order, the RGE equation for $\lambda$ is
influenced both by the top Yukawa coupling and the electroweak gauge couplings.
The full RGE for $\lambda$, rather than Eq. (11), reads~\cite{Hpoles}
\begin{equation}
\frac{d\lambda(\mu)}{d\ln\mu^2} = \frac{3}{4\pi^2}
\left[\lambda^2(\mu)-\frac{1}{4}\lambda_t^4(\mu) + \frac{\pi}{128}
[3+2\sin^2\theta_W + \sin^4\theta_W]\alpha_2(\mu^2)\right]~.
\end{equation}
The important point to notice in this equation is the {\bf negative}
contribution coming from the top coupling.  This contribution, just like the
$\alpha_3$ contribution in Eq. (55), can cause $\lambda(\mu)$ to
{\bf decrease} at first.  Indeed, if the Higgs coupling $\lambda(M_H)$ is
not large enough, because the Higgs boson is light, the relatively large
contribution coming from the $\lambda_t^4$ term can drive $\lambda(\mu)$
{\bf negative} at some scale $\mu$.  This cannot really happen physically,
because for $\lambda < 0$ the Higgs potential is unbounded!

To avoid this vacuum instability below some cut-off $\Lambda_c$--typically
$\Lambda_c\sim M_{\rm P}$--one needs to have $\lambda(M_H)$, and therefore
the Higgs mass, sufficiently large.  Hence, these considerations 
give a lower bound for the
Higgs mass.  Taking $\Lambda_c = M_{\rm P}$, this 
lower bound is~\cite{lowerbound}
\begin{equation}
M_H \geq 134~{\rm GeV}~.
\end{equation}
Lowering the cut-off $\Lambda_c$, weakens the bound
on $M_H$.  Interestingly, to have a SM Higgs as light as 100 GeV---which
is the region accessible to LEP 200---requires a very low cut-off,
of order $\Lambda_c\sim 100$ TeV.~\cite{Lindner}  So, finding such a Higgs may, very
indirectly, point to a scenario with extra compact dimensions where such
low-cut-offs are allowed.  Parenthetically, I should note that these kinds
of vacuum stability bounds cease to be valid in models with more than one
Higgs doublet, like in the SUSY SM case.

\subsubsection{SUSY SM}

The situation is quite different for $\lambda_t$ if there is 
supersymmetric matter.  Because supersymmetry necessitates two Higgs doublets,
$\lambda_t(m_t)$ is no longer fixed solely by the value of the top mass.
The vacuum expectation values of the two Higgs bosons involve a further
parameter, $\tan\beta$, besides $v_F$:
\begin{equation}
\langle H_u\rangle = \frac{1}{\sqrt{2}}v_F\sin\beta~; ~~~~
\langle H_d\rangle = \frac{1}{\sqrt{2}} v_F\cos\beta~.
\end{equation}
Thus, one has, instead of Eq. (54),
\begin{equation}
\lambda_t(m_t) = \frac{0.95\pm 0.03}{\sin\beta}~.
\end{equation}

Keeping again only the leading terms, the RGE for $\lambda_t(\mu)$ in the
presence of SUSY matter reads now:
\begin{equation}
\frac{d\lambda_t(\mu)}{d\ln\mu^2} = \frac{1}{32\pi^2}
\left[6\lambda_t^2(\mu)+\lambda_b^2(\mu)-\frac{64\pi}{3}
\alpha_3(\mu^2)\right]\lambda_t(\mu)~.
\end{equation}
Because of the differrent coefficients that SUSY matter implies, it is no 
longer necessarily true that the square bracket above is negative at
$\mu = m_t$, as was the case in the SM.  Instead, since
$64\pi\alpha_3(m_t^2)/3 \simeq 7.9$, the square bracket above can actually
vanishes in two regions of parameter space.  
The first is the region of large $\tan\beta$
where, for large scales $\mu$,
$\lambda_b(\mu)\simeq \lambda_t(\mu)\simeq 1$ [Yukawa unification
region].  The second is a region where $\tan\beta\sim O(1)$, so that
$\lambda_b(m_t)\ll \lambda_t(m_t)$, with the top contribution cancelling
directly that coming from the $SU(3)$ corrections [in detail, this requires
$\sin\beta = 0.83\pm 0.03$].

In either of the two regions above, the presence of SUSY matter forces
$\lambda_t$ to an infrared fixed point~\cite{Hill} as $\mu$ 
becomes of order $m_t$
\begin{equation}
\left.\frac{d\lambda_t(\mu)}{d\ln\mu^2}\right|_{\mu\sim m_t} \simeq 0~.
\end{equation}
This is a very interesting possibility, since such a condition essentially
serves to drive quite different values of $\lambda_t(\mu)$ at high scales
$\mu$ down to the same fixed point value $\lambda^*$, at scales
$\mu\simeq m_t$.  This behavior is illustrated in Fig. 4.  
Asking that Eq. (63) holds, one
sees that the fixed point $\lambda^*$ is given by
\begin{equation}
\lambda^* = \frac{32\pi}{9}\alpha_3(m_t^2)\left[
\frac{1}{1+\frac{m_b^2(m_t)\tan^2\beta}{6m_t^2(m_t)}}\right] \simeq 1.3
\left[\frac{1}{1+\frac{m_b^2(m_t^2)\tan^2\beta}{6m_t^2(m_t)}}\right]~.
\end{equation}

\begin{figure}[t]
\begin{center}
\epsfig{file=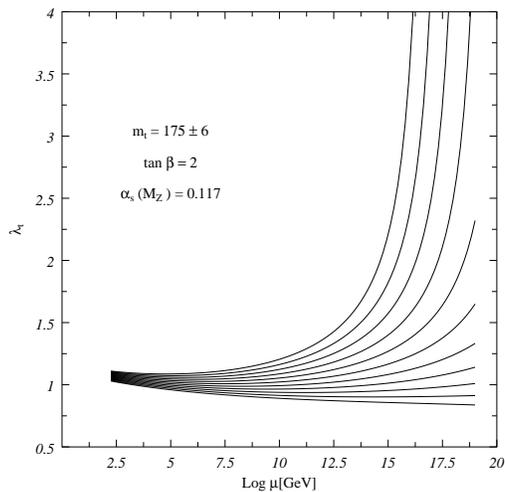,height=3in}
\end{center}
\caption{Focusing of the Yukawa couplings as $\lambda_t\to\lambda^*$.}
\end{figure}

Two remarks are in order
\begin{description}
\item{i)} The fixed point behavior for $\lambda_t(m_t)$ does, indeed,
need supersymmetry.  In the SM with only ordinary matter, one would get a
fixed point behavior from the RGE for $\lambda_t$ only if $m_t$ would have
been approximately 250 GeV!
\item{ii)} If one were to assume that near the Planck mass $\lambda_t(\mu)$
were to be large then, if there is SUSY matter, the fixed point behavior for
$\lambda_t$ essentially serves to predict the correct value for the top mass
$m_t(m_t)\simeq m_t^*\simeq 170$ GeV seen by experiment.
\end{description}

\subsection{Neutrino Oscillations}

Although hints of neutrino oscillations have been around for some time,
notably connected with the solar neutrino puzzle,\cite{Bahcall} real evidence
for oscillatioins only emerged last year from data on atmospheric neutrinos
studied by the large underground water Cerenkov detector, SuperKamiokande.
In June 1998, the SuperKamiokande Collaboration~\cite{SK} reported a
pronounced zenith angle dependence for the flux of multi-GeV atmospheric
$\nu_\mu$ events, but no such dependence for the atmospheric $\nu_e$ flux.
The collaboration interpreted the large up-down asymmetry seen
[139 up-going $\nu_\mu$'s versus 256 down-going $\nu_\mu$'s] as evidence
for $\nu_\mu\to\nu_X$ oscillations, with $\nu_X$ being either a $\nu_\tau$ or,
possibly, a new sterile neutrino $\nu_s$.\footnote{A sterile neutrino is one
that has no $SU(2)\times U(1)$ interactions.  Because $\nu_s$ does not have
any couplings to the $Z$-bosons it does not contribute to the $Z$ width, so
a light $\nu_s$ is not excluded by the precise neutrino counting result from LEP: $N_\nu = 2.991\pm 0.011$.\cite{LEPEW}}

In the usual 2-neutrino mixing formalism,\cite{RDPMex} the weak interaction
eigenstates $|\nu_\mu\rangle$ and $|\nu_X\rangle$ are linear combinations
of two mass eigenstates $|\nu_1\rangle$ and $|\nu_2\rangle$:
\begin{equation}
\left(\begin{array}{l}
|\nu_\mu\rangle \\ |\nu_X\rangle
\end{array} \right) = \left(
\begin{array}{ll}
\cos\theta & \sin\theta \\ -\sin\theta & \cos\theta
\end{array} \right) \left(
\begin{array}{l}
|\nu_1\rangle \\ |\nu_2\rangle
\end{array} \right)~.
\end{equation}
The probability that after traversing a distance $L$ a $\nu_\mu$ neutrino of
energy $E$ emerges as a $\nu_X$ neutrino is then given by the well known
formula
\begin{equation}
P(\nu_\mu\to\nu_X;L) = \sin^22\theta~\sin^2
\frac{\Delta m^2L}{4E}~,
\end{equation}
with $\Delta m^2=m_2^2-m_1^2$ being the difference in mass squared between
the two neutrino eigenstates.  From an analysis of their results, the
SuperKamiokande collaboration~\cite{SK} deduced that the observed
up-down asymmetry could be explained by neutrino oscillations if the mixing
was nearly maximal, $\sin^22\theta\simeq 1$, and if  $\Delta m^2\simeq
2.5\times 10^{-3}~{\rm eV}^2$.

The SuperKamiokande results provide a {\bf lower bound} on neutrino masses.
Since
\begin{equation}
\Delta m^2 = m_2^2-m_1^2\simeq 2.5\times 10^{-3}~{\rm eV}^2~,
\end{equation}
it follows that at least one neutrino has a mass larger than
\begin{equation}
m_2 \geq 5\times 10^{-2}~{\rm eV}~,
\end{equation}
with the bound being satisfied if $m_2\gg m_1$.  Such small masses, compared
to the masses of quarks and leptons, suggests that new physics is at work.
Indeed, the simplest way to understand why neutrinos 
have tiny masses is through the see-saw mechanism,\cite{GRSY} which involves
new physics at a scale much larger than the Fermi scale.  Let me briefly
discuss this reasoning.

Because neutrinos have no charge, as we remarked earlier, 
the most general mass term for these states
can contain also particle-particle terms, besides the usual particle-antiparticle
contribution.~\cite{RDPMex} One has, considering one species of neutrinos for
simplicity,
\begin{eqnarray}
{\cal{L}}_{\rm mass}^\nu &=& -m_D[\bar\nu_{\rm L}\nu_{\rm R} +
\bar\nu_{\rm R}\nu_{\rm L}] - \frac{m_{\rm R}}{2}
[\bar\nu_{\rm R}\tilde C\bar\nu_{\rm R}^T + \nu_{\rm R}^T\tilde C
\nu_{\rm R}] \nonumber \\
& & -\frac{m_{\rm L}}{2}[\nu_{\rm L}^T\tilde C\nu_{\rm L} + 
\bar\nu_{\rm L}\tilde C\bar\nu_{\rm L}^T]~. 
\end{eqnarray}
In the above the different mass terms conserve/violate different
symmetries.  To wit:
\begin{description}
\item{} The Dirac mass $m_D$:  conserves lepton number L, but violates
$SU(2)\times U(1)$.
\item{} The Majorana mass $m_{\rm R}$:  violates lepton number L, but
conserves \break $SU(2)\times U(1)$.
\item{} The Majorana mass $m_{\rm L}$:  violates both lepton number L and
$SU(2)\times U(1)$.
\end{description}

Because the Dirac mass term has the same form as the usual quark and lepton
masses, it is sensible to imagine that $m_D$ should be of the same order of
magnitude as these masses.  Hence, schematically, one expects 
\begin{equation}
m_D\sim m_\ell\sim v_F
\end{equation}
where the proportionality constant to the Fermi scale may, indeed, be quite small.
Thus, if one wants the physical neutrinos to have very  small masses, this must
be the result of the presence of the Majorana mass terms in Eq. (69).

There are two simple ways of achieving this goal, depending
on whether one assumes that a right-handed neutrino $\nu_{\rm R}$ exists or not.
If one does not involve a $\nu_{\rm R}$, the simplest effective interaction
one can write using only $\nu_{\rm L}$ which preserves $SU(2)\times U(1)$
is~\cite{Weinberg}
\begin{equation}
{\cal{L}}_{\rm eff} = \frac{1}{M} (\nu_\ell~\ell)_{\rm L}^T 
\tilde C\vec\tau \left(
\begin{array}{c}
\nu_\ell \\ \ell 
\end{array} \right)_{\rm L} \cdot
\Phi^T C\vec\tau\Phi + {\rm h.c.}
\end{equation}
Here $M$ is some, presumably large, scale which is associated with these
lepton number violating processes.  This term, when $SU(2)\times U(1)$
breaks down to $U(1)_{\rm em}$, generates a mass for the neutrino
\begin{equation}
m_\nu \equiv m_{\rm L} = \frac{v_F^2}{M}~.
\end{equation}
One sees that to get neutrino masses of the order of those inferred from 
SuperKamiokande one requires $M\sim 10^{15}$ GeV---a scale of the order of the
GUT scale!

One can get a similar result if one includes right-handed neutrinos in the
theory.  In this case, it is convenient to rewrite the general neutrino
mass terms of Eq. (69) in terms of both neutrino fields, $\nu$, and their
charged conjugate, $\nu^c$.  Since~\cite{RDPMex}
\begin{equation}
\nu^c = \tilde C\bar\nu^T~; ~~~  \overline{\nu^c} = \nu^T\tilde C~,
\end{equation}
Eq. (69) takes the form
\begin{equation}
{\cal{L}}^\nu_{\rm mass} = -\frac{1}{2}\left[\left(\overline{(\nu_{\rm L})^c}~
\overline{\nu_{\rm R})}\right) \right]\left(
\begin{array}{cc}
m_{\rm L} & m_D \\ m_D & m_{\rm R}
\end{array} \right) \left(
\begin{array}{c}
\nu_{\rm L} \\ {(\nu_{\rm R})^c} 
\end{array} \right)  
+ {\rm h.c.} ~.
\end{equation}

If one neglects $m_{\rm L}$ altogether $(m_{\rm L}=0)$ and $m_{\rm R} \gg
m_D$, then it is easy to see that the eigenvalues of the neutrino mass matrix
\begin{equation}
{\cal{M}} = \left(
\begin{array}{cc}
0 & m_D \\ m_D & m_{\rm R}
\end{array} \right)
\end{equation}
have a large splitting, producing a heavy neutrino and one ultralight
neutrino:
\begin{equation}
m_{\rm heavy} \sim m_{\rm R}~; ~~~~
m_{\rm light} \sim \frac{m_D^2}{m_{\rm R}}~.
\end{equation}
This is the famous see-saw mechanism.\cite{GRSY}

In the see-saw mechanism the light neutrino state, $n$, is mostly $\nu_{\rm L}$,
while the heavy neutrino state, $N$, is mostly $\nu_{\rm R}$:
\begin{eqnarray}
\nu_{\rm L} &\simeq & n_{\rm L} + \frac{m_D}{m_{\rm R}} N_{\rm L} \nonumber \\
\nu_{\rm R} &\simeq & N_{\rm R} - \frac{m_D}{m_{\rm R}} n_{\rm R}~.
\end{eqnarray}

Assuming that     
SuperKamiokande has observed $\nu_\mu\to\nu_\tau$ oscillations,
and that the light neutrino has a mass given by Eq. (76), then
\begin{equation}
m_\nu = \frac{m_D^2}{m_{\rm R}} \sim 5\times 10^{-2}~{\rm eV}~.
\end{equation}
If $m_D\simeq m_\tau$, one requires $m_{\rm R}\sim 10^{11}~{\rm GeV}$.
If $m_D\simeq m_t$, on the other hand, one requires $m_{\rm R}\sim 
10^{15}$ GeV.  Irrespective of the choice, again one sees that to obtain neutrino
masses in the sub-eV range via the see-saw mechanism one needs to involve
new scales, connected to $m_{\rm R}$, which are much above $v_F$.

These two examples make it clear that the neutrino oscillations detected by
SuperKamiokande are definitely signs of new physics.  However, it is quite
likely that this new physics is disconnected from the precise mechanism which
causes the $SU(2)\times U(1)_{\rm em}$ breakdown.  This is certainly the
case if the light neutrinos involved in the oscillations are produced
by the see-saw mechanism, since the parameter $m_{\rm R}$ is an
$SU(2)\times U(1)$ singlet.  Thus the scale $m_{\rm R}$ has nothing at all to
do with $v_F$.  This is likely to be true also if the light neutrinos are
generated by effective interactions of the type shown in Eq. (71).  Although
light neutrino masses in this case arise only after $SU(2)\times U(1)$
breaking, the physics that gives origin to these masses is the physics
associated with the scale $M$ (likely some GUT physics) 
characterizing the effective interactions.

Because of the above, neutrino oscillations are unlikely to give much
information on the nature of the physics which gives rise to the Fermi
scale $v_F$.  For this reason,  
in what follows, I shall not pursue this interesting topic
further, prefering to concentrate instead on the dynamics of electroweak
symmetry breaking.

\section{Promises and Challenges of Dynamical Symmetry Breaking}

The idea behind a dynamical origin for the Fermi scale $v_F$ is rather
simple.\cite{SW}  One imagines that there exists an underlying strong
interaction theory which confines and that the fundamental fermions $F$
of this theory carry also $SU(2)\times U(1)$ quantum numbers.  If the
confining forces acting on $F$ allow the formation of $\langle\bar FF\rangle$
condensates then, in general, these condensates will cause the breakdown of
$SU(2)\times U(1)$, since $\bar FF$ also carries non-trivial 
$SU(2)\times U(1)$ quantum numbers.  The dynamical scale $\Lambda_F$
associated with the underlying strongly interacting theory is then,
{\it de facto}, the Fermi scale:
\begin{equation}
\langle\bar FF\rangle \sim \Lambda_F^3\sim v_F^3~.
\end{equation}

There are two generic predictions of such theories:
\begin{description}
\item{i)} Because of the strongly coupled nature of the underlying theory,
there should be no light Higgs boson in the spectrum.
\item{ii)} Just like in QCD, this underlying theory should have a rich
spectrum of bound states which are singlets under the symmetry group of the
underlying theory.  These states, typically, should have masses
\begin{equation}
M\sim \Lambda_F\sim v_F~.
\end{equation}
Among these states there should be a heavy Higgs boson.
\end{description}

It has become conventional to denote the underlying strong interaction theory
responsible for the breakdown of $SU(2)\times U(1)\to U(1)_{\rm em}$
as Technicolor, which was the name originally used by Susskind.\cite{SW}
Just as there are two generic predictions for Technicolor theories, there are
also two necessary requirements for these theories coming from experiment.
These are
\begin{description}
\item{iii)} The underlying theory must lead naturally to the connection between
$W$ and $Z$ masses embodied in the statement that $\rho = 1$, up to radiative
corrections.  That is $M_W^2 = M_Z^2\cos^2\theta_W$.  
As we shall see, this obtains if the underlying theory has some, protective, $SU(2)$ global
symmetry.
\item{iv)} The Technicolor spectrum must be such that the parameter $\epsilon_3$
defined in Section 3 is small, as seen experimentally.  For this to be so,
one needs that
\begin{equation}
\int^\infty_0 \frac{ds}{s}[R_V(s)-R_A(s)] < 0~.
\end{equation}
\end{description}
The first requirement above is easy to achieve in most Technicolor theories 
(but, eventually, quite constraining).
The second requirement is much 
harder to implement, since it requires understanding some unknown
strong dynamics!

Let me begin by discussing how one can 
guarantee that the underlying theory give
$\rho = 1$.  For this purpose, it proves useful to examine how this happens
in the SM.  There $\rho = 1$ emerges as a result of an {\bf accidental
symmetry} in the Higgs potential.  If one writes out the complex Higgs
doublet $\Phi$ in terms of real fields
\begin{equation}
\Phi = \frac{1}{\sqrt{2}}\left(
\begin{array}{c}
\phi_1 + i\phi_2 \\ \phi_3 + i\phi_4
\end{array} \right)~,
\end{equation}
it is immediately clear that the potential
\begin{equation}
V = \lambda\left(\Phi^\dagger\Phi - \frac{v_F^2}{2}\right)^2
\end{equation}
has a bigger symmetry than $SU(2)\times U(1)$, namely $O(4)$.  The VEV of
$\Phi$ is, in the notation of Eq. (82), the result of $\phi_1$ getting a
VEV:  $\langle \phi_1\rangle = v_F$.  Obviously, this VEV causes the
breakdown of $O(4)\to O(3)$.  It is the remaining $O(3)$ symmetry, after
the spontaneous breakdown, which forces $\rho=1$.  Indeed, this symmetry
requires the 11, 22 and 33 matrix elements of the weak-boson mass
matrix to have all the same value:
\begin{equation}
M^2 = \frac{1}{4} v_F^2 \left[
\begin{array}{cccc}
g_2^2 & 0 & 0 & 0 \\ 0 & g_2^2 & 0 & 0 \\
0 & 0 & g_2^2 & g_2g^\prime \\ 0 & 0 & g_2g^\prime & g^{\prime 2}
\end{array} \right]~,
\end{equation}
giving $\rho = 1$.

To guarantee $\rho = 1$ in Technicolor models one must build in the
same custodial $O(3)\sim SU(2)$ symmetry present in the Higgs potential.
Such a custodial $SU(2)$ symmetry in fact exists in QCD with just 2 flavors.
Neglecting the $u$- and $d$-quark masses the QCD Lagrangian has an
$SU(2)_{\rm L}\times SU(2)_{\rm R}\times U(1)_{\rm L+R}$ global 
symmetry.\footnote{The $U(1)_{\rm L-R}$ symmetry present at the Lagrangian
level is not preserved at the quantum level,
because of the nature of the QCD vacuum.\cite{RDP1}}  However, only the vectorial piece of the
$SU(2)_{\rm L}\times SU(2)_{\rm R}\sim SU(2)_V\times SU(2)_A$ symmetry
survives as a good symmetry (Isospin) of QCD, since the formation of 
$\langle\bar uu\rangle = \langle\bar dd\rangle\not= 0$ condensates breaks the
$SU(2)_A$ symmetry spontaneously.

Given the circumstances described above, the simplest way to guarantee that
$\rho = 1$ in Technicolor models is to make these models look very much
like QCD.  Indeed, this was the strategy adopted originally by
Susskind and Weinberg.~\cite{SW}  One organizes the underlying Technifermions
$F$ in doublets
\begin{equation}
\left(
\begin{array}{c}
U_i \\ D_i
\end{array} \right) ~~~~
i = 1,\ldots,N_D
\end{equation}
but assumes, just like in QCD, that  
the left- and right-handed components of these states transform
differently under $SU(2)\times U(1)$:
\begin{equation}
\left(
\begin{array}{c}
U_i \\ D_i
\end{array} \right)_{\rm L} \sim 2~; ~~~~
U_{i{\rm R}}\sim 1;~ D_{i{\rm R}}\sim 1~.
\end{equation}
Neglecting the electroweak interactions, if the Technifermions are massless,
then the Technicolor theory has a large global chiral symmetry
$SU(2N_D)_{\rm L}\times SU(2N_D)_{\rm R} \supset
SU(2)_{\rm L}\times SU(2)_{\rm R}$.  The condensates which one
assumes form due to the strong Technicolor forces, and which break
$SU(2)\times U(1)\to U(1)_{\rm em}$,
\begin{equation}
\langle\bar U_{i{\rm L}}U_{i{\rm R}}\rangle =
\langle\bar D_{i{\rm L}}D_{i{\rm R}}\rangle \not= 0~,
\end{equation}
also break this global symmetry down.  In particular, 
$SU(2)_{\rm L}\times SU(2)_{\rm R}\to SU(2)_{\rm L+R}$ and this
custodial $SU(2)$ symmetry serves to guarantee that in the gauge
boson spectrum $M^2_W = M^2_Z\cos^2\theta_W$.

If one pushes the QCD-Technicolor analogy a bit more, one can infer
something about the scale of the Technicolor mass spectrum.  Both QCD and
Technicolor have an approximate $SU(2)_{\rm L}\times SU(2)_{\rm R}$
global symmetry broken down to $SU(2)_{\rm L+R}$.  For QCD such a
breakdown gives rise to the pions, $\vec\pi$, as Nambu-Goldstone bosons.
Technicolor, analogously will have also three Technipions $\vec\pi_T$.
However, in contrast to the pions which are real states, the Technipions,
when $SU(2)\times U(1)$ is gauged, become the longitudinal components of the
$W^\pm$ and $Z$ gauge bosons.  Nevertheless, from the $\vec\pi-\vec\pi_T$
analogy one can estimate the importance of Technicolor interactions and
their associated spectra.  More specifically $\pi-\pi$ scattering in QCD
should tell us something about $\pi_T-\pi_T$ scattering in Technicolor.

For $\pi-\pi$ scattering one can write a partial wave expansion for the
scattering amplitude ${\cal{A}}$ of the form
\begin{equation}
{\cal{A}} = 32\pi \sum_J (2\pi+1) P_J (\cos\theta) a_J(s)~.
\end{equation}
The $SU(2)\times SU(2)$ chiral symmetry of QCD allows one to calculate the
$s$-wave scattering amplitudes $a_o$ at threshold, and one finds~\cite{pipi}
\begin{equation}
a_o^{I=0} = \frac{s}{16\pi f_\pi^2}~; ~~~~
a_o^{I=2} = -\frac{s}{32\pi f_\pi^2}
\end{equation}
where $f_\pi$ is the pion decay constant.  Unitarity requires the 
partial wave amplitudes to be bounded $|a_J|\leq 1$, with strong interactions
signalled by these amplitudes saturating the bound.  Using Eq. (89), naively
one sees that
this occurs when the energy squared $s\sim 16\pi f_\pi^2$.  Indeed, at these
energies $\pi-\pi$ scattering already is dominated by resonance formation.
For the Technicolor theory, by analogy, one should have similar formulas
with $f_\pi$ replaced by $v_F$.  Hence, if the analogy 
holds, one should expect Technicolor resonances to appear at an 
energy scale of order
$4\sqrt{\pi} v_F \sim 1.7$ TeV.  If one trusts this estimate, the physics
of the underlying Technicolor theory will be hard to see, even at the
LHC!

The analogy between a possible Technicolor theory and QCD, however, cannot be
pushed too far.  Indeed, we know from our discussion of the precision
eletroweak tests that the spectrum of vector and axial resonances in the
Technicolor theory must be quite {\bf different} than QCD, since what is
required is that Eq. (81) hold---which has the {\bf opposite} sign of what
obtains in QCD!  Thus Technicolor, in some 
fundamental aspects, must be quite different
than QCD.  This also emerges from a 
different set of considerations, connected with 
the mass spectrum of quarks and leptons.  I turn to this issue now. 

Although strange, and difficult to implement in practice, it is possible to
imagine that a Technicolor theory exists in which, as far as the electroweak
radiative corrections go, the presence of a heavy Higgs state around a TeV and
a Technicolor spectrum in the few TeV range combine to mimic the effects of
a single light $(M_H\sim 100~{\rm GeV})$ Higgs, giving a tiny $\epsilon_3$.  
A much harder task is
to ask that this theory also generate a {\bf realistic} mass spectrum for
the quarks and leptons.  In my view, this latter problem is the principal
difficulty of Technicolor theories.  

In weakly coupled theories, where the
Fermi scale is a parameter put in by hand, one can easily generate quark and
lepton masses through Yukawa couplings.  In these theories there is no
reason that the physics which is associated with $v_F$ be connected 
to the physics which served to 
produce the Yukawa couplings.  Indeed, it is likely that this
latter physics  
is one associated with scales much larger than $v_F$.
This freedom of decoupling the origin of the quark and lepton mass spectrum
from the Fermi scale does not exist for theories where $v_F$ is generated
through a strongly coupled theory.  
In these theories one is forced to try to understand
fermion mass generation at scales of order of the Fermi scale $v_F$, or just
one or two orders of magnitude higher.  This complicates life immensely.

To generate quark and lepton masses in Technicolor theories at
all, one must introduce some communication between these states---which I
shall denote collectively as $f$---and the Technifermions $F$, whose
condensates cause the electroweak breakdown.  This necessitates, in general,
introducing yet {\bf another} strongly coupled underlying theory, which has
been dubbed extended Technicolor (ETC).~\cite{DSEL}  Spontaneous ETC
breakdown, in conjunction with Technicolor-induced electroweak breakdown,
is at the root of the quark and lepton mass spectrum.  
However, since at least one state
in this spectrum, top, has a mass of $O(v_F)$, the scale associated with the
ETC breakdown cannot be very much larger than $v_F$.
Let me discuss this in a bit more detail.  

The ETC interactions couple
the ordinary fermions $f$ to the Technifermions $F$.  
As a result, the exchange of an
ETC gauge boson between pairs of $fF$ states, when ETC spontaneously breaks
down, generates an effective interaction which, schematically, reads
\begin{equation}
{\cal{L}}^{\rm ETC}_{\rm eff} = \frac{1}{\Lambda^2_{\rm ETC}}
\bar F_{\rm L}F_{\rm R}\bar f_{\rm R}f_{\rm L} + {\rm h.c.}
\end{equation}
Such an interaction generates a mass term for the ordinary fermions, as the
result of the formation of the $SU(2)\times U(1)$ breaking Technifermion
condensate $\langle\bar F_{\rm L}F_{\rm R}\rangle\sim v_F^3$.  Thus one
finds
\begin{equation}
m_f = \frac{\langle\bar F_{\rm L}F_{\rm R}\rangle}{\Lambda^2_{\rm ETC}}
\sim\frac{v_F^3}{\Lambda^2_{\rm ETC}}~.
\end{equation}
If one tries to use this formula to produce a top mass, since $m_t$ itself
is of $O(v_F)$ one sees, as alluded to above, that the ETC scale $\Lambda_{\rm ETC}$ cannot be large---typically, $\Lambda_{\rm ETC}\sim (1-10)$ TeV.
Such a low ETC scale, however, is troublesome because it generally leads to too large flavor changing neutral currents (FCNC).  For instance, as discussed long
ago by Dimopoulos and Ellis,\cite{DE} the box graph containing
both ETC and Technifermion exchanges gives a contribution to the 
$\bar K^o-K^o$ mass difference which is far above that coming from the
weak interactions, unless $\Lambda_{\rm ETC} \stackrel{>}{_{\scriptstyle
\sim}} 100$ TeV.

Although the FCNC $\leftrightarrow m_t$ conundrum is not the only problem of
Technicolor/ETC models,\footnote{For instance, because of the large global
symmetries present in the Technicolor sector, condensate formation leads to the
appearance of many more (pseudo) Nambu-Goldstone bosons than just the
Technipions,
$\vec\pi_T$.  Unless other interactions can generate sufficiently large masses
for these extra 
states, their presence in the Technicolor spectrum
invalidates these theories, since no trace
of these states 
has yet been seen experimentally.~\cite{Treview}} its theoretical
amelioration has been a principal target for partisans of dynamical symmetry
breaking.~\cite{Chivukula}  Fortunately, as I will discuss below, one of the
more interesting ``solutions"---dubbed Walking Technicolor (WTC)~\cite{Holdom}---involves theories that are rather different from QCD
dynamically.  It is perhaps not unreasonable to hope that for such theories
the constraint (81) actually might hold.  Some arguments in favor of this
contention actually exist.~\cite{sundrum}

The essence of how WTC models ameliorate the $m_t\leftrightarrow$ FCNC
conundrum can be readily appreciated by noting that the Fermi scale $v_F$
and the Technifermion condensate $\langle\bar FF\rangle$, in general, are
sensitive to rather {\bf different} parts of the self-energy of Technifermions.
Since $v_F$ measures the strength of the matrix element of the spontaneously
broken currents between a Technipion state and the vacuum, $v_F^2$ is
proportional to an integral over the square of the Technifermion self-energy.
Schematically,
this result, gives for $v_F^2$ the formula~\cite{Pagels}
\begin{equation}
v_F^2\sim\int\frac{d^4p}{(p^2)^2} \Sigma^2(p)\sim\Sigma^2(0)~,
\end{equation}
where the second approximation is valid up to logarithmic terms.  The condensate
$\langle\bar FF\rangle$, on the other hand, just involves an integral over
the Technifermion self-energy.  Again, schematically,
\begin{equation}
\langle\bar FF\rangle \sim \int\frac{d^4p}{p^2}\Sigma(p)\sim
\Lambda^2_{\rm ETC} \Sigma(\Lambda_{\rm ETC})~.
\end{equation}
The second term above, again to logarithmic accuracy, recognizes that this
integral is dominated by the largest scales in the theory.  In our case, 
since one integrates up to the ETC scale, this is $\Lambda_{\rm ETC}$.

One can understand how WTC theories work from Eqs. (92) and (93), augmented
by a result of Lane and Politzer~\cite{LP} detailing the asymptotic behavior
of fermionic self-energies in theories with broken global symmetries.
Lane and Politzer showed that
\begin{equation}
\Sigma(q^2) \stackrel{q^2\to\infty}{\sim}
\frac{\Sigma^3(0)}{q^2} \sim\frac{\Lambda^3}{q^2},
\end{equation}
where the last line recognizes that the self-energy $\Sigma(0)$
scales according to 
the dynamical scale $\Lambda$ of the theory in question.  For
ordinary Technicolor theories, by the time one has reached the ETC scale,
the Technifermion self-energy should have already reached the asymptotic
form (94).  Thus
\begin{equation}
\langle\bar FF\rangle\sim \Lambda^2_{\rm ETC}
\Sigma(\Lambda_{\rm ETC}) \sim\Sigma^3(0)
\sim v_F^3
\end{equation}
and the condensate $\langle\bar FF\rangle$ indeed scales as $v_F^3$, as
we have assumed.  This, however, is not the case for WTC theories.  In these
theories, the evolution with $q^2$ of the WTC coupling constant, as well as
of the Technifermion self-energy, is very slow--hence the name, Walking
Technicolor.  In particular, the Technifermion self-energy at the ETC scale
is assumed to be nowhere near the asymptotic form (94) so that,
\begin{equation}
\Sigma(\Lambda_{\rm ETC}) \gg \frac{\Lambda^3_{\rm TC}}{\Lambda^2_{\rm ETC}}~.
\end{equation}
In this case, there is a large disparity between the condensate 
$\langle\bar FF\rangle$ and $v_F^3$:
\begin{equation}
\langle\bar FF\rangle\sim \Lambda^2_{\rm ETC}\Sigma(\Lambda_{\rm ETC})
\gg \Lambda^3_{\rm TC} \sim v_F^3~.
\end{equation}

Although Walking Technicolor theories are quite interesting dynamically,
and some WTC theories have been constructed which are semirealistic,~\cite{semi} there are still many difficulties in practice,
notably with top itself.  For instance, to really get the top mass
large enough, one has to really have very slow ``walking".  Since we want
$m_t\sim v_F$ and $m_t$ is given by the formula
\begin{equation}
m_t\sim \frac{\langle\bar FF\rangle}{\Lambda^2_{\rm ETC}} \sim
\Sigma(\Lambda_{\rm ETC})
\end{equation}
one needs
\begin{equation}
\Sigma(\Lambda_{\rm ETC}) \sim \Sigma(0)~.
\end{equation}
Such ``slow walking" again is only realistic if $\Lambda_{\rm ETC}$ is
relatively near to $v_F$.  
However, then FCNC problems re-emerge, even in the
WTC context!  Furthermore, the large ETC effects that are used to boost the
top mass up cause other problems.  In particular, as Chivukula, Selipsky and Simmons pointed out,~\cite{CS} the same graphs that give rise to the top mass produce
a rather large anomalous $Zb\bar b$ vertex.  In the simplest WTC model, this
gives an unacceptably large shift for the ratio $R_b$---the ratio of the
rate of $Z\to b\bar b$ to that of hadrons---$(\delta R_b)_{\rm ETC}\simeq -0.01$.

To avoid the problems alluded to above, as well as other 
problems,~\cite{Chivukula} lately some
hybrid models have been developed.  These, so called, topcolor-Technicolor
models~\cite{tt} get the masses for all first and second generation quarks 
and leptons from a WTC/ETC theory.  However, the top (and bottom) masses
come from yet a 
{\bf third} underlying strong interaction theory--topcolor ~\cite{topcolor}--which produces
4-Fermi effective interactions involving these states.  So top, effectively,
gets its mass from the presence of a top-quark condensate 
$\langle\bar tt\rangle$,
formed as a result of the topcolor theory.  Because $\langle\bar tt\rangle$
also breaks $SU(2)\times U(1)$, in these topcolor-Technicolor models, the
Fermi scale arises both as the result of these condensates and of the usual
Technifermion condensates $\langle\bar FF\rangle$.  

Although these theories
have some attractive features, and some interesting predictions of
anomalies in top interactions,~\cite{Hill2} one has moved a long way away from
the simple idea that the electroweak breakdown is much like the BCS theory
of superconductivity!  Of course, ultimately,  
only experiment will tell if these rather complicated
ideas have merit or not.  From the point of view of simplicity, however, the
weak coupling SM alternative of invoking the existence of low energy
supersymmetry seems a more desirable route to follow.  I turn to this topic
next.

\section{The SUSY Alternative}

The procedure for constructing a supersymmetric extension of the SM is
straightforward, and rather well known by now.~\cite{SUSYreview}  For 
completeness, let me outline the principal steps here.  They are:
\begin{description}
\item{i)} One associates scalar partners to the quarks and leptons
(squarks and sleptons) and fermion partners to the Higgs scalar(s)
(Higgsinos), building chiral supermultiplets
\begin{equation}
{\bf \{\Psi\}} = \{\tilde\psi,\psi\}~,
\end{equation}
composed of complex scalars $\tilde\psi$ and Weyl fermions $\psi$.  For
instance the left-handed electron chiral supermultiplet
\begin{equation}
{\bf \{\Psi_e\}} = \{\tilde e_{\rm L},e_{\rm L}\}
\end{equation}
contains a (left-handed) selectron $\tilde e_{\rm L}$ and a left-handed
electron.
\item{ii)} One associates spin-1/2 partners to the gauge fields (gauginos),
building vector supermultiplets
\begin{equation}
{\bf \{V\}} = \{V^\mu,\lambda\}~,
\end{equation}
composed of a gauge field $V^\mu$ and a Weyl fermion gaugino $\lambda$.
\item{iii)} One supersymmetrizes all interactions.  For example, the ordinary Yukawa coupling of the Higgs $H_d$ to the
quark doublet $Q_{\rm L}$ and the quark singlet $d_{\rm R}$ is now 
accompanied by two other vertices  
involving $\tilde H_d~Q_{\rm L}~\tilde d_{\rm R}$
and $\tilde H_d~\tilde Q_{\rm L}~d_{\rm R}$.\footnote{Here $\tilde H_d$ is
the spin-1/2 SUSY partner of $H_d$ and $\tilde Q_{\rm L},\tilde d_{\rm R}$
are the spin-0 partners of $Q_{\rm L},~d_{\rm R}$, respectively.}
\end{description}

In addition to the above three points, supersymmetry imposes some
constraints on which kind of interactions are allowed.\cite{SUSYreview}
For our purposes here, two of these constraints are most significant:
\begin{description}
\item{iv)} Interactions among chiral superfields are derivable from a
{\bf superpotential} $W({\bf \{\Phi_i\}})$ which involves only
${\bf \{\Phi_i\}}$ and not ${\bf \{\Phi_i^*\}}$.
\item{v)} The scalar potential $V(\phi_i)$ follows directly from the
superpotential, plus terms---the, so called, $D$-terms---arising from gauge
interactions~\cite{SUSYreview}
\begin{equation}
V(\phi_i) = \sum_i\left|\frac{\partial W}{\partial\phi_i}\right|^2 +
\frac{g_a}{2}\left|\phi_i^*t_a\phi_i\right|^2~.
\end{equation}
Here $t_a$ is the appropriate generator matrix for the scalar field $\phi_i$
for the symmetry group whose coupling is $g_a$.
\end{description}

I remark that point (iv) above is the reason one needs two different Higgs
fields in supersymmetry.  Even though the scalar field $H_d^*$ has the
same quantum numbers as $H_u$, a superpotential term $W_3$
(${\bf H_d^*}$, ${\bf Q_{\rm L}}$, ${\bf u_{\rm R}}$) is not
allowed.  There is another way to understand why a second Higgs
supermultiplet is needed in a supersymmetric extension of the SM, involving
anomalies.  The Higgsino $\tilde H_d$ has opposite charges to the Higgsino
$\tilde H_u$.  Omitting one of these two fields in the theory would
engender a chiral anomaly in the $SU(2)\times U(1)$ theory, since for
$SU(2)\times U(1)$ to be anomaly free it is necessary that~\cite{anomalies}
\begin{equation}
{\rm Tr}~[Q]_{\rm fermions} = 0~.
\end{equation}
Although Eq. (104) holds for the quarks and leptons, it would fail if
only $\tilde H_d$ was included and not $\tilde H_u$.  So anomaly 
consistency requires two Higgs bosons in a SUSY extension of the SM.

With this brief precis of the SUSY SM in hand, let me examine what are the
implications of supersymmetry 
for the issue of $SU(2)\times U(1)$ breaking.  If one ignores
for the moment any Yukawa couplings, the only superpotential term one is
allowed to write down for the SUSY SM is
\begin{equation}
W({\bf H_u},{\bf H_d}) = \mu {\bf H_uH_d}~.
\end{equation}
This superpotential gives the following scalar potential for the SUSY SM
[cf. Eq. (103)]
\begin{eqnarray}
V(H_u,H_d) &=& \mu^2[H_u^\dagger H_u + H_d^\dagger H_d] + 
\frac{1}{8} g^{\prime 2}
[H_u^\dagger H_u - H_d^\dagger H_d]^2 \nonumber \\
& & +\frac{1}{8} g_2^2[H_u^\dagger\vec\tau H_u + H_d^\dagger\vec\tau
H_d]\cdot [H_u^\dagger\vec\tau H_u + H_d^\dagger\vec\tau H_d]~.
\end{eqnarray}
However, because all the coefficients in the above are positive, 
it is clear that
the potential $V$ {\bf cannot} break $SU(2)\times U(1)$!

This is really not a disaster, since a SUSY SM is not a realistic theory
without including some breaking of the supersymmetry.  Once one adds some
SUSY breaking terms to Eq. (106), then it is quite possible that these
terms can cause the $SU(2)\times U(1)\to U(1)_{\rm em}$ breakdown.
However, not any type of supersymmetry breaking terms are allowed.  To
preserve the solution to the hierarchy problem which supersymmetry provided
(namely, logarithmic sensitivity to the cutoff---$\ln\Lambda_c$---and not
quadratic sensitivity---$\Lambda_c^2$) one must ask that the SUSY-breaking
Lagrangian has {\bf only} terms of dimensionally $d<4$.  Thus, schematically,
${\cal{L}}_{\stackrel{\rm SUSY}{\rm breaking}}$ 
has the form~\cite{SUSYbreaking}
\begin{equation}
{\cal{L}}_{\stackrel{\rm SUSY}{\rm breaking}} = 
-\sum_{ij} \mu^2_{oij}\phi^\dagger_i\phi_j -
\sum _im_i\lambda_i\lambda_i + \sum_jA_jW_{3j}(\phi_i) + BW_2(\phi_i)~.
\end{equation}
In the above the $\phi_i$ fields are scalars and the $\lambda_i$ fields
are gauginos.  The coefficients $\mu^2_{oij}$ are scalar mass terms for the
$\phi_i,\phi_j$ scalars; $m_i$ are possible gaugino masses; and $A_i$ and
$B$ are scalar coefficients which multiply the trilinear and bilinear 
couplings of scalar fields which follow from the form of the
superpotential.

Including SUSY breaking terms, the Higgs potential is modified to
\begin{eqnarray}
V(H_u,H_d) &=& (H_u^\dagger~ H_d^\dagger){\cal{M}}^2 \left(
\begin{array}{c}
H_u \\ H_d
\end{array} \right) +
\frac{1}{8} g^{\prime 2}[H_u^\dagger H_u-H_d^\dagger H_d]^2 \nonumber \\
& & +\frac{1}{8} g_2^2[H_u^\dagger\vec\tau H_u + H_d^\dagger
\vec\tau H_d]\cdot[H_u^\dagger\vec\tau H_u + H_d^\dagger\vec\tau H_d]~,
\end{eqnarray}
where the mass squared ${\cal{M}}^2$ is given by
\begin{equation}
{\cal{M}}^2 = \left(
\begin{array}{cc}
\mu^2 + \mu_{11}^2 & -B\mu + \mu_{12}^2 \\
-B\mu + \mu_{12}^2 & \mu^2 + \mu^2_{22} 
\end{array} \right) \equiv \left(
\begin{array}{cc}
m_1^2 & m_3^2 \\ m_3^2 & m_2^2 
\end{array} \right)~.
\end{equation}
Obviously, a breakdown of $SU(2)\times U(1)\to U(1)_{\rm em}$ is now
possible {\bf provided}
\begin{equation}
{\rm det}~{\cal{M}}^2 < 0~.
\end{equation}
Note that the Higgs potential in the SUSY SM, even though it involves 2 Higgs
fields, is considerably more restricted than that of the SM.  In particular,
the quartic terms of the potential are {\bf not} arbitrary.  Because they 
arise from the $D$-terms, these couplings are fixed by the strength of the
gauge interactions themselves.

If $SU(2)\times U(1)\to U(1)_{\rm em}$, the spectrum of the potential $V$ of
Eq. (108) contains 5 physical scalars: $h;H;A;$ and $H^\pm$.  The first three
of these are neutrals, with two scalars $h$ and $H$ and one pseudoscalar 
$A$.\footnote{By convention $M_h<M_H$.}  The masses of all 5 of these states
are functions of the 5 independent parameters entering in $V$, namely
$(m_1^2,m_2^2,m_3^2,g^{\prime 2},g^2_2)$ or, more physically, the set of three
masses ($M^2_W,M^2_Z,M_A^2$) and two mixing angles ($\tan\beta,\cos\theta_W$).
However, since $M^2_W=M^2_Z\cos^2\theta_W$, because of doublet Higgs
breaking, and because $M_W$ and $M_Z$ are well measured experimentally, in effect the Higgs
spectrum in the SUSY SM has only two unknowns: $\tan\beta$ and $M_A^2$.

A straightforward calculation, using the potential of Eq. (108) yields the
following results~\cite{Higgsguide}
\begin{eqnarray}
M^2_{H^\pm} &=& M^2_A + M^2_W \nonumber \\
M^2_{H,h} &=& \frac{1}{2} (M_A^2 + M^2_Z) \pm \frac{1}{2}
\left[(M_A^2 + M^2_Z)^2 - 4M_Z^2 M_A^2\cos^22\beta\right]^{1/2}~.
\end{eqnarray}
It is easy to see from Eq. (111) that there is always one {\bf light Higgs}
in the spectrum:
\begin{equation}
M_h \leq M_Z|\cos 2\beta| \leq M_Z~.
\end{equation}
However, the bound of Eq. (112) is not trustworthy, as it is quite sensitive
to radiative effects which are enhanced by the large top mass.~\cite{Higgsred}
Fortunately, the magnitude of the radiative shifts for $M^2_h$ can be well
estimated, by either direct conputation~\cite{Hollik} or via the
renormalization group.~\cite{Haber}

It is useful to illustrate the nature of these radiative shifts for $M_h^2$
using the RGE in the special limit in which $M_h\to (M_H)_{\rm SM}$, but where
$(M_H)_{\rm SM}$ is fixed to be at the $Z$-mass.  This limit~\cite{Haber}
obtains as $M_A$ gets very large $(M_A\to\infty)$ and $|\cos 2\beta|\to 1$.
In the above limit, the only remaining light field in the theory
is $M_h$ and, since $M_h=M_Z$ at tree level, the theory is just like the SM
except that the quartic Higgs coupling $\lambda$, in the SM, is fixed to
\begin{equation}
\lambda_{\rm SM} = \frac{1}{8} (g_2^2+g^{\prime 2})~.
\end{equation}
Recalling the RGE [Eq. (58)] for the evolution of $\lambda$, with its large
{\bf negative} contribution due to $\lambda_t$, one sees that, approximately
\begin{equation}
\lambda(\mu) = \lambda(M_h) - \frac{3}{8\pi^2} \lambda_t^4
\ln \frac{\mu}{M_h}~.
\end{equation}

Using Eq. (114) one can readily estimate the radiative shift from the tree
level equation $M_h = M_Z$.  One has
\begin{equation}
M_h^2 = 2\lambda(M_h) v_F^2 = 2\lambda(\mu) v^2_F +
\frac{3}{8\pi^2} \lambda_t^4 \ln\frac{\mu}{M_h}~.
\end{equation}
Using Eq. (113)
and taking $\mu$ to be the characteristic scale of the SUSY partners- 
$\tilde m$- 
the, relatively, slow running of the gauge couplings allows one to write,
approximately
\begin{equation}
2\lambda(\tilde m) v^2_F = \frac{1}{4} \left.(g_2^2 + g^{\prime 2})\right|_{\tilde m} v^2_F\simeq \frac{1}{4} \left.(g_2^2 + g^{\prime 2})\right|_{M_Z} v_F^2=
M_Z^2~.
\end{equation}
Hence Eq. (115) yields the formula~\cite{Haber}
\begin{equation}
M_h^2 = M_Z^2 + \frac{3\alpha m_t^4}{2\pi\sin^2\theta_W M^2_W}
\ln\frac{\tilde m^2}{M^2_Z}~.
\end{equation}
This is quite a large
shift since, for the scale of the
SUSY partners $\tilde m\simeq 1$ TeV, numerically one finds
$\Delta M_h\simeq 20$ GeV.

Eq. (117) was obtained in a particular limit $(|\cos 2\beta|\to 1)$, but
an analogous result can be obtained for all $\tan\beta$.  It turns out that
for small $\tan\beta$
the shifts are even larger than those indicated in Eq. (117). However, for these values of $\tan\beta$ the tree order contribution is also
smaller, since $\left.M_h\right|_{\rm tree} < M_Z\cos 2\beta$.  
To illustrate this point, the expectations
for $M_h$, plotted as a function of $\tilde m$ is shown in Fig. 5  for two values
of $\tan\beta$.  

\begin{figure}[t]
\begin{center}
\epsfig{file=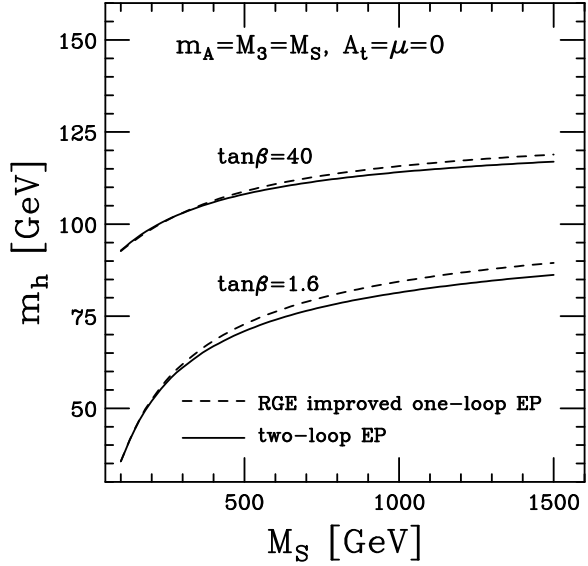,height=3in}
\end{center}
\caption{Plots of $M_h$ as a function of $\tilde m\equiv M_S$ for two values of $\tan \beta$, from Ref 56.}
\end{figure}

The results shown in this figure~\cite{Zhang} neglect any
details in the SUSY spectrum, 
since they have all been subsumed in the average parameter
$\tilde m$.  It turns out that the most important effect of
the SUSY spectrum for $\Delta M_h$ 
arises if there is an incomplete cancellation between
the top and the stop contributions, due to large $\tilde t_{\rm L}-\tilde t_{\rm R}$ mixing.~\cite{Haber}  At their maximum these effects
can cause a further shift of order $(\Delta M_h)_{\rm mixing} \simeq 10$ GeV.

One can contrast these predictions of the SUSY SM with experiment.  At LEP 200,
the four LEP collaborations have looked both for the process 
$e^+e^-\to hZ$ and $e^+e^-\to hA$.  The first process is analogous to that
used for searching for the SM Higgs, while $hA$ production is peculiar
to models with two (or more) Higgs doublets.  One can show that these two
processes are complementary, with one dominating in a region of parameter 
space where the other is small, and {\it vice versa}.~\cite{Higgsguide}
LEP 200 has established already rather strong bounds for $M_h$ and $M_A$
setting the 95\% C.L. bounds (for $\tan\beta > 0.8$)~\cite{HVan}
\begin{equation}
M_h > 77~{\rm GeV}~; ~~~~ M_A > 78~{\rm GeV}
\end{equation}
As Fig. 6 shows, if there is not much $\tilde t_{\rm L}-\tilde t_{\rm R}$
mixing the low $\tan\beta$ region $[0.8 < \tan\beta < 2.1]$ is also already
excluded.

\begin{figure}
\begin{center}
\epsfig{file=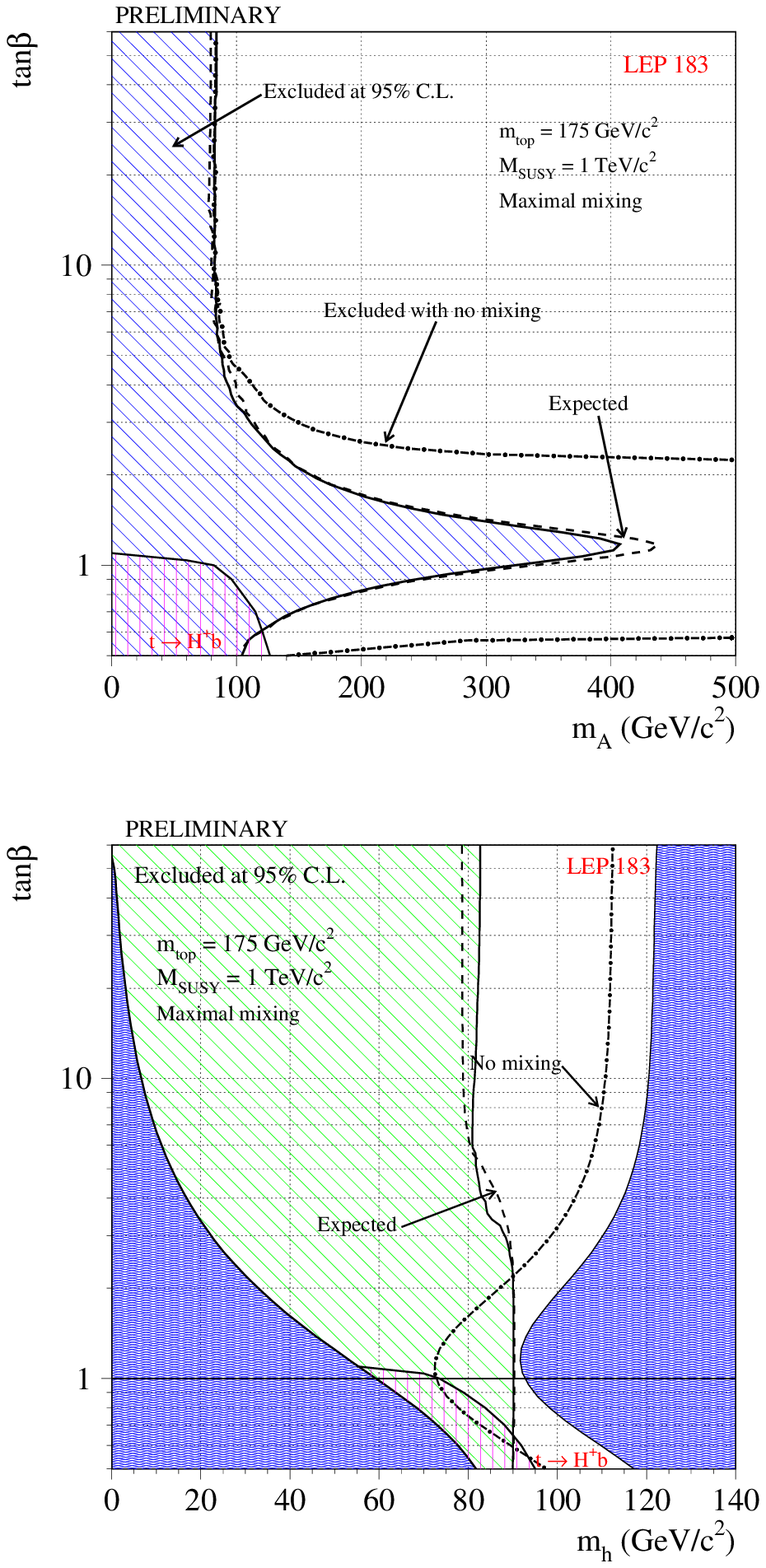,height=6in}
\end{center}
\caption{LEP200 limits for$M_h$ and $M_A$ as a function of $\tan \beta$, from Ref. 57.}
\end{figure}

Although the SUSY SM is rather predictive when it comes to the Higgs sector,
beyond this sector the spectrum of SUSY partners and possible allowed
interactions is quite model dependent.  Most supersymmetric extensions of the
SM considered are assumed to contain a discrete symmetry, $R$-parity, which
is conserved.  This assumption simplifies considerably the form of the
possible interactions one has to consider.  In fact, $R$-parity
conservation provides an essentially unique way to generalize the SM since
$R$, defined by
\begin{equation}
R = (-1)^{Q+L+2J}~,
\end{equation}
with $Q$ being the quark number, $L$ the lepton number and $J$ the spin,
turns out simply to be +1 for all particles and -1 for all sparticles.

Obviously, $R$ parity conservation implies that SUSY particles enter in
vertices always in pairs, and hence sparticles are always pair produced.
This last fact implies, in turn, the {\bf stability} of the lightest
supersymmetric particle (LSP), even in the presence of supersymmetry
breaking interactions.  Although supersymmetry must be broken, since we do
not observe multiplets of particles and sparticles of the same mass, SUSY
breaking interactions are quite restrictive and do not end up by violating
the stability of the LSP.  Let me discuss the issue of SUSY breaking in a
little more detail, since the manner in which one breaks supersymmetry is the
principal source of model-dependence for the SUSY SM.

In general,~\cite{Peskin2} one assumes that SUSY is spontaneously broken at
some scale $\Lambda$ in some {\bf hidden sector} of the theory.  This sector is
coupled to ordinary matter by some {\bf messenger} states of mass $M$, with
$M \gg \Lambda$, and all that obtains in the visible sector is a set of soft
SUSY breaking terms---terms of dimension $d<4$ in the Lagrangian of the 
theory.\footnote{Terms of $d=4$ would re-introduce the hierarchy problem.}
Ordinary matter contains supersymmetric states with masses $\tilde m\sim$ TeV,
with $\tilde m$ given generically by
\begin{equation}
\tilde m\sim \frac{\Lambda^2}{M}~.
\end{equation}

Within this general framework, two distinct scenarios have been suggested
which differ in what one assumes are the messengers that connect the hidden
SUSY breaking sector with the visible sector.  In supergravity models~\cite{SUGRA} (SUGRA), the messengers are gravitational interactions,
so that $M\sim M_{\rm Planck}$.  Then, because of Eq. (120) and the demand
that $\tilde m\sim$ TeV, the scale of SUSY breaking in the hidden sector is
of order $\Lambda\sim 10^{11}$ GeV.  In contrast, in models where the
messengers are gauge interactions ~\cite{GMM} (Gauge Mediated Models) with
$M\sim 10^6$ GeV, then the scale of spontaneous breaking of supersymmetry is
around $\Lambda\sim 10^3$ TeV.

In both cases one assumes that the supersymmetry is a local symmetry, gauged
by gravity.~\cite{Ferrara}  Then the massless fermion which originates from
spontaneous SUSY breaking, the goldstino, is absorbed and serves to give
mass to the spin-3/2 gravitino---the SUSY partner of the graviton.
This mass is of order
\begin{equation}
m_{3/2} \sim \frac{\Lambda^2}{M_{\rm P}}~.
\end{equation}
Obviously, in SUGRA models the gravitino has a mass of the same order as
all the other SUSY partners ($\tilde m \sim$ TeV).  
However, in SUGRA models, in general, one
does not assume that the gravitino is the LSP.  However, in Gauge
Mediated Models, since $\Lambda \ll 10^{11}$ GeV, the gravitino
is definitely the LSP.

Besides the above difference, the other principal difference between SUGRA
and Gauge Mediated Models of supersymmetry breaking is the assumed form of
the soft breaking terms.  In SUGRA models, to avoid FCNC problems, one
needs to assume that the soft breaking terms are {\bf universal}.  This
assumption is unnecessary in Gauge Mediated Models, where in fact one can
explicitly compute the form of the soft breaking terms and show that they do not
lead to FCNC.  Let me discuss this
a little further.

The basic point is the following.  As I alluded to earlier [cf. Eq. (107)],   as a result of supersymmetry breaking one
ends up, in general, with nondiagonal mass terms for the scalar fields
$\phi_i$:
\begin{equation}
{\cal{L}}_{\stackrel{\rm soft}{\rm mass}} = -\sum_{ij}
\mu^2_{oij}\phi_i^\dagger\phi_j~.
\end{equation}
These terms, which connect states of the same charge, can lead to large FCNC
through effective mass diagonal couplings of gluinos to squarks and quarks.
The scalar mass insertion (122) gives rise to a $\tilde g d_L\tilde s_L$ vertex. Such a vertex can generate a very large $K^o-\bar K^o$ mixing
term, through a gluino-squark box graph. 
With SUSY masses in the TeV range, this would be a total disaster.
Hence, effectively, only diagonal soft mass terms can be
countenanced.  In the SUGRA models, this circumstance forces one to consider
only {\bf universal} soft breaking terms, with
\begin{equation}
\mu^2_{oij} = \mu^2_o \delta_{ij}~.
\end{equation}

In the case of Gauge Mediated Models, this flavor blindness arises more
naturally since SUSY breaking is the result of gauge interactions which
are flavor diagonal.  As a result, the soft mass terms can only couple diagonally
and the soft masses will be proportional to the gauge couplings squared.
One can show that the soft mass for the $i^{\rm th}$ scalar field is given by~\cite{GMM}
\begin{equation}
\mu_i^2 = \sum^3_{a=1} \frac{\alpha_a^2}{(4\pi)^2} C_i
\frac{\Lambda^4}{M^2}~,
\end{equation}
where $C_i$ is the appropriate Casimir factor for the scalar field in
question.  It follows from this equation that in Gauge Mediated Models, in
general, squarks are heavier than sleptons since the latter do not have
strong interactions.  In SUGRA models both squarks and sleptons are
assumed to have the same mass at large scales $(\mu\stackrel{>}{_{\scriptstyle \sim}} \Lambda)$.  However, as a result of the evolution of couplings these
universal masses can become quite different at scales of order 100 GeV.
Thus SUGRA models at low scales turn out to be not so dissimilar in their spectrum to Gauge Mediated Models.

This point is particularly germane for gauginos, where the differences between
SUGRA and Gauge Mediated Models are quite small.  For Gauge Mediated
Models, the analogous equation to Eq. (124) for gaugino masses is only 
quadratically dependent on the gauge couplings~\cite{GMM}
\begin{equation}
m_a = \frac{\alpha_a}{4\pi} \frac{\Lambda^2}{M}~.
\end{equation}
Hence the ratio of the $SU(3)$, $SU(2)$ and $U(1)$ gaugino   
masses scale with the $\alpha_i$
\begin{equation}
m_1:m_2:m_3 = \alpha_1:\alpha_2:\alpha_3~.
\end{equation}
In SUGRA models, although at high scales one assumes a universal gaugino
mass $m_{1/2}$, the masses of the individual gauginos evolve in the same
way as the squared gauge coupling constants
\begin{equation}
\frac{dm_i}{d\ln\mu^2} = -\frac{1}{16\pi^2} b_im_i~.
\end{equation}
Whence, at low scales, Eq. (126) holds again.  As a result, the most important
difference between these two SUSY breaking scenarios is that in Gauge
Mediated Models the gravitino is the LSP, while in SUGRA models, the LSP is,
in general, thought to be a neutralino---a spin-1/2 partner to the neutral
gauge and Higgs bosons.

In Gauge Mediated Models, because the gravitino is the LSP, an important role
for phenomenology is played by the ``next lightest" SUSY state---the NLSP.
In general, the NLSP in these models is either a slepton or a neutralino~\cite{GMM}
and, because it is not the lightest, this NLSP is unstable.  The decay
\begin{equation}
{\rm NLSP} \to \mbox{ordinary state + gravitino}
\end{equation}
has a lifetime which scales as
\begin{equation}
\tau_{\rm NLSP} \to \frac{\Lambda^4}{M^5_{\rm NLSP}}~.
\end{equation}
Depending on whether the NLSP decays occur within, or outside, the detector,
the phenomenology of the Gauge Mediated Models will be similar, or rather
different, to that of SUGRA models. 
This is because the gravitino acts essentially as
missing energy, which is the signal associated with the SUGRA LSP.

I will not explicitly discuss here what are 
the expected signals for either the SUGRA
or the Gauge Mediated scenarios, since the expected phenomenology is both
involved and quite dependent on the actual spectrum of supersymmetric 
states assumed.~\cite{SUSYrev}  I note here only that the present Tevatron and LEP 200
lower bounds on SUSY-states, although somewhat model dependent, are in the
neighborhood of 100 GeV.  More precisely, the weakest bounds are for the
neutralinos $(M_{\chi^o}\stackrel{>}{_{\scriptstyle \sim}} 30$ GeV), followed
by charginos $(M_{\chi^\pm}\stackrel{>}{_{\scriptstyle \sim}} 70$ GeV),
sleptons $(M_{\tilde\ell}\stackrel{>}{_{\scriptstyle \sim}} 90$ GeV) and
squarks and gluinos $(m_{\tilde q},m_{\tilde g}\stackrel{>}{_{\scriptstyle
\sim}} 200$ GeV).~\cite{SUSYrev}

Before closing this Section, I would like to make two final points concerning
the SUSY alternative.  These are
\begin{description}
\item{i)} Not only does supersymmetry allow for a stable hierarchy between
the Planck mass and the Fermi scale, but supersymmetry breaking itself can
help trigger the $SU(1)\times U(1)\to U(1)_{\rm em}$ breaking.  The Higgs
mass matrix ${\cal{M}}^2$ can start with ${\rm det}~ {\cal{M}}^2>0$ at high scales but
can evolve to ${\rm det}~{\cal{M}}^2<0$ at the Fermi scale.  Indeed $m_2^2$, the mass
term associated with the $H_u$ Higgs, is strongly affected by the large
top Yukawa coupling and can change sign as one evolves to low $q^2$.  This
radiative breaking of $SU(2)\times U(1)$, triggered by SUSY breaking,~\cite{Ross}
is a very attractive feature of the SUSY SM.  For it to produce the Fermi
scale, however, one must inject already at very high scales a SUSY breaking
mass parameter also of order $v_F$.  Why this should be so remains
a mystery.
\item{ii)} In SUGRA models the neutralino LSP, which is a linear combination
of all the neutral weak gaugino and Higgsino fields
\begin{equation}
\tilde\chi = \alpha_\gamma\tilde\gamma + \alpha_Z\tilde Z + 
\alpha_u\tilde H_u + \alpha_d\tilde H_d
\end{equation}
is an excellent candidate for cold dark matter.~\cite{Dettino}  
In the Gauge
Mediated case, the gravitino could provide a warm dark matter candidate,~\cite{Masiero} provided that 
\begin{equation}
m_{3/2} \sim\frac{\Lambda^2}{M_{\rm P}} \sim {\rm KeV}~,
\end{equation}
which requires $\Lambda\sim 10^3$ TeV.  The presence of possible candidates 
for the dark matter in the Universe, is another quite remarkable feature
of low energy SUSY models and speaks in favor of the SUSY alternative.
\end{description}

\section{Are there Extra Dimensions of Size Greater than a (TeV)$^{-1}$?}

To conclude these lectures, I want to make some remarks on the phenomenological
consequences of imagining that the Fermi scale is associated with the Planck
scale of $(d+4)$-dimensional gravity.  That is
\begin{equation}
v_F \sim (M_{\rm P})_{d+4}\equiv M~.
\end{equation}
The extra $d$-dimensions in this theory are assumed to be compact and of size
$R$.  By comparing the Newtonian forces in $(d+4)$-dimensions with that in
four dimensions, one can interrelate $M$ and the usual Planck mass
$M_{\rm P} = G_N^{-1/2} \simeq 10^{19}$ GeV.~\cite{ADD}  One has
\begin{equation}
F_4 = \frac{1}{M_{\rm P}^2} \frac{m_1m_2}{r^2}~; ~~~
F_{4+d} = \frac{1}{(M)^{2+d}} \frac{m_1m_2}{r^{d+2}}~.
\end{equation}
Then, using Gauss' law, it follows that~\cite{ADD}
\begin{equation}
M_{\rm P} = M(MR)^{d/2}~.
\end{equation}

Obviously, demanding that $M$ be of the order of the Fermi scale $v_F$, fixes
the size $R$ of the compact dimensions as a function of the number of these
extra dimensions $d$.  One finds that for $d=2$, $R\sim 10^{-1}$ cm, while 
for $d=6$, $R\simeq 10^{-13}$ cm.  For distances $r$ below the size $R$ of the
compact dimensions one will begin to register the fact that gravitational
forces do not follow the familiar $r^{-2}$ law.  Thus the first test that this
idea is not nonsensical is to look for possible modifications of the usual
Newtonian potential at short distances.  For $r\sim R$, the potential will
feel Yukawa-like correctione of strength $2d$:~\cite{PP}
\begin{equation}
V(r) = -\frac{G_Nm_1m_2}{r}\left\{1+2de^{-r/R} + \ldots\right\}~.
\end{equation}

There have been a variety of tests of the Newtonian potential at (relatively)
short distances.  For $r>100\mu$, the most stringent limits for deviation
from the Newtonian expectations~\cite{HMP} allow $R\sim 10^{-1}$ cm for 
strengths $\alpha\equiv 2d$ of $O(1)$.  Below $100\mu$, the best bounds on
non-Newtonian forces come from Casimir experiments but, as shown in Fig. 7, these experiments only put some constraints on modifications which have a
strength $\alpha$ much below unity. This figure also shows the limits which
might be achievable in a proposed experiment employing 1 KHz mechanical oscillators as test masses.  This experiment~\cite{HMP}
should be able to push $\alpha$ to $\alpha\sim 10^{-1}$ for compact
dimensions $R\sim 10^{-1}$ cm.  Thus it could {\bf directly
test} for deviations from Newtonian gravity to a level where an effect might be seen.

\begin{figure}
\begin{center}
\epsfig{file=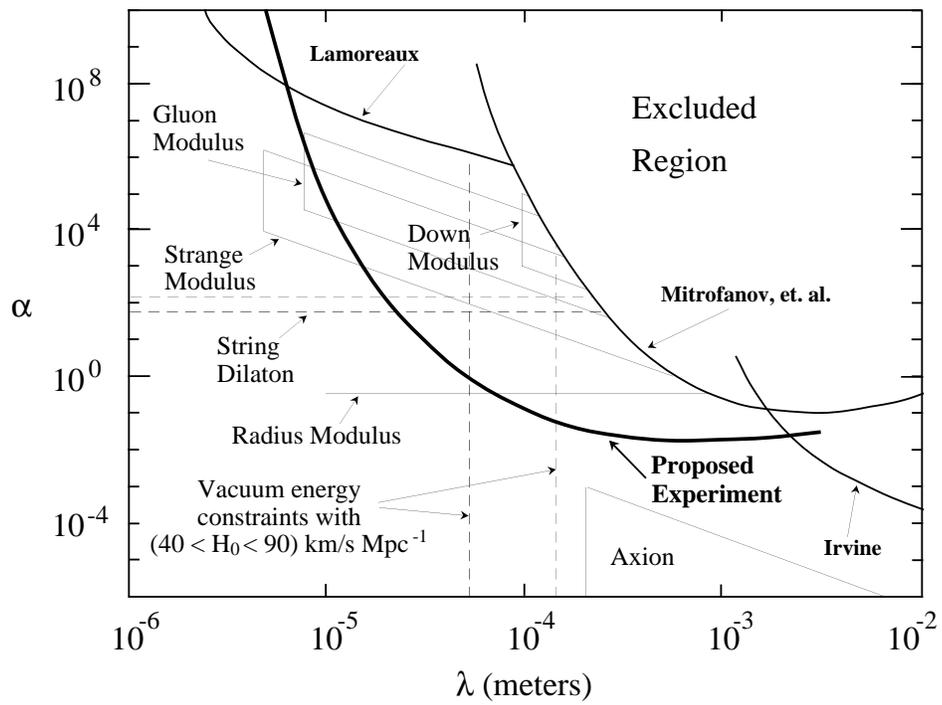,height=5.5in}
\end{center}
\caption{Experimental status of gravitational strength forces at short distances, from Ref. 67.}
\end{figure}

At high energy, the presence of extra compact dimensions of finite size $R$ can be detected because these theories allow abundant  graviton production to occur. Obviously, to see any effects one needs experiments at
energies $(\sqrt{s})_{\rm parton}\sim M \sim v_F$-- energies which will be
attainable at the LHC.  The presence of 
compact dimensions of size $R$ allows gravitons to get produced, at energies of order
$(\sqrt{s})_{\rm parton} \sim M$, with a probability proportional to
$M^{-2}$ {\bf not} $M_{\rm P}^{-2}$.  However, the graviton production
is rather soft, with the softness being greatest the larger the number of
compact dimensions $d$ is.

This can be understood qualitatively as follows.  In the theories under
discussion, the SM fields exist essentially on a 4-dimensional hypersurface,
with gravity acting both on this hypersurface, as well as on a set of
compact dimensions of size $R$.  From the point of view of the 4-dimensional
theory, a graviton with its momentum $p^\alpha$ acting in the compact
dimensions is equivalent to having a particle of mass
\begin{equation}
m_{\rm eff}\sim\frac{i}{R}~,
\end{equation}
where $i$ is an integer associated with the level of excitation.
It follows that, for  given total energy $E$, one can access many states in
the compact dimensions.  Using (136), the spacing among these states is of
order $\Delta E\sim R^{-1}$.  Thus, for processes of total energy, $E$, 
the total number of states probed in
the compact dimensions is
\begin{equation}
N\sim\left(\frac{E}{\Delta E}\right)^d \sim (RE)^d~.
\end{equation}

The probability of gravitational production at $(\sqrt{s})_{\rm parton} = E$
is inversely proportional to the Planck mass $M_{\rm P}$, but directly 
proportional to the number of states $N$ excited in the compact dimensions.
If $N$ is large, which will occur for $E\gg R^{-1}$, then, effectively, the
probability of producing gravitational radiation is much greater than the 
classical expectations.  Indeed, as alluded to above, 
this probability scales like $M^{-2}$ not
$M_{\rm P}^{-2}$.  One has
\begin{equation}
{\rm Probability} \sim \frac{1}{M_{\rm P}^2} N \sim
\frac{(ER)^d}{M_{\rm P}^2} = \frac{(ER)^d}{M^{2+d} R^d} =
\frac{1}{M^2}\left(\frac{s}{M^2}\right)^{d/2}~.
\end{equation}
Note, however, that even though the probability indeed scales like $M^{-2}$
there is an additional soft infrared multiplying factor of $(s/M^2)^{d/2}$.
Therefore, one learns that strong graviton production comes on slowly for
$s<M^2$.  Thus the importance of these effects for the LHC, even if these
theories were to be true, is crucially dependent on the value of $M$.

Mirabelli, Perelstein and Peskin,~\cite{PP} as well as others,~\cite{others}
have studied graviton production at LEP and Tevatron energies to try to
obtain bounds on $M$ from present day data.  These authors have looked for
the processes $e^+e^-\to \gamma G$ and $p\bar p\to {\rm jet}~G$, with the graviton
experimentally being manifested as missing energy.  Analyses of the LEP data
for the process $e^+e^-\to\gamma\nu\bar\nu$ and of the CDF/DO bounds for
monojet production yield the bounds:
\begin{equation}
 M\stackrel{>}{_{\scriptstyle \sim}} 600 {\rm GeV}~~(d=6);~~~~
M\stackrel{>}{_{\scriptstyle \sim}} 750{\rm GeV}~~(d=2).
\end{equation}

Obviously the LHC will be sensitive to much greater values of $M$.  However,
at least for $d=2$, astrophysics already puts a strong constraint on $M$.
Basically, if $M$ is too low graviton emission will cool off supernovas too
quickly.  The fact that SN 1987a does not show any anomalous cooling then
allows one to set a bound of $M\stackrel{>}{_{\scriptstyle \sim}} 30$ TeV,
for $d=2$.~\cite{ADD2}

\section{Concluding Remarks}

In these lectures I have discussed the theoretical arguments for believing
that there is physics beyond the electroweak theory, related to the origin and
magnitude of the Fermi scale $v_F$.  Each of the suggestions for new physics
which I examined: dynamical symmetry breaking, supersymmetry and extra
compact dimensions, had its own characteristic signals and its own set of
theoretical puzzles to resolve.  {\it A priori}, it is clearly not possible to
select among these alternatives purely theoretically. However, from the few meager experimental hints for physics beyond the SM that we have today,  supersymmetry does appear to be favored.  What is exciting
is that we should know quite soon which of
these alternatives, if any, are correct. If one is lucky, the answer may come perhaps already from experiments at LEP200
or the Tevatron. However, the origin of the Fermi scale should certainly become clear once the LHC starts running.

\section*{Acknowledgements}

I am grateful to Bruce Campbell and Fakir Khanna for the wonderful hospitality at Chateau Lake Louise.  This work was supported in part by the Department of Energy under contract No. DE-FG03-91ER40662, Task C.

\end{document}